# Perceptions of AI Across Sectors:
# A Comparative Review of Public Attitudes

Filip Biały, Mark Elliot and Robert Meckin
The University of Manchester

**Abstract**

This paper offers a domain-mediated comparative review of 251 studies on public attitudes toward AI, published between 2011 and 2025. Drawing on a systematic literature review, we analyse how different factors including perceived benefits and concerns (or risks) shape public acceptance of - or resistance to - artificial intelligence across domains and use-cases, including healthcare, education, security, public administration, generative AI, and autonomous vehicles. The analysis highlights recurring patterns in individual, contextual, and technical factors influencing perception, while also tracing variations in institutional trust, perceived fairness, and ethical concerns. We show that the public perception in AI is shaped not only by technical design or performance but also by sector-specific considerations as well as imaginaries, cultural narratives, and historical legacies. This comparative approach offers a foundation for developing more tailored and context-sensitive strategies for responsible AI governance.

## 1. Introduction

Even though current generation of AI is underpinned by a common technology - namely machine learning, especially in the form of deep learning - in the public eye it has not emerged as a single solution. Rather, it has taken shape through multiple and overlapping applications - ranging from predictive diagnostics in healthcare and algorithmic hiring systems in HR to autonomous weapons and generative language models. As AI becomes increasingly embedded in sector-specific infrastructures, the question of how publics perceive its use is gaining urgency.

Existing literature on public perception of AI suggests that attitudes are highly sensitive to the application domain. People tend to be more supportive of AI in domains where it is perceived to augment human capacity (e.g., in medical diagnostics) and more sceptical when AI is seen as replacing judgement or threatening civil liberties or rights (e.g., in security or surveillance). These perceptions are shaped not only by technical features of the AI system but also by institutional trust, cultural attitudes toward risk, and the moral economy of the domain in question. Despite this, few reviews have systematically compared public perceptions across sectors and explored the cross-domain patterns and differences in attitudes.

This paper addresses that gap. Through a review of 251 academic and grey literature sources, we examine how public perceptions of AI differ across several major domains, such as healthcare, education, security, autonomous transportation, and public administration, as well as across different types or forms AI takes in these and other domains, including generative AI, automated decision-making, and robotics. For each, we explore what drives acceptance, what triggers resistance, and how different social groups relate to AI in context. At the same time, we will identify existing gaps and inconsistencies in the way in which public perception of AI has been studied and offer suggestions on how to make such research more robust.

Our guiding questions are:

- RQ1: How do public attitudes toward AI differ across domains of application, and what accounts for these differences?
- RQ2: What are the perceived benefits and concerns in each domain, and how are they influenced by individual (demographic), contextual, and technical factors?
- RQ3: To what extent do institutional trust, transparency, and human oversight mediate public acceptance of AI technologies?
- RQ4: What lessons can be drawn for sector-specific governance and public engagement strategies?

By adopting a comparative, domain-sensitive lens, this paper aims to support a more granular understanding of public attitudes and to inform more equitable and effective governance of AI.

In what follows, we describe the literature review methodology, including the search and screening strategy. In the third section we outline our theoretical assumptions and explain the key concepts and analytical framework that utilises three types of factors (individual, contextual, and technical) in tracing the cross-domain patterns of AI perceptions. In the fourth section we present the results of the review. We start with the overview of the dataset and papers that survey general (non-domain-specific) perceptions and trust in AI and move to domain-specific studies. We then synthesise the findings by describing the patterns in line with the three categories of factors influencing perception of AI and by offering a typology of cross-domain as well as domain-specific perceived benefits and concerns related to AI. In the fifth section we discuss the findings, and limitations of this study. We conclude with suggestions on directions of future research.

## 2. Methodology

### Literature review methodology

This working paper is based on a systematic review of academic and grey literature on public perceptions of artificial intelligence. The initial search was conducted in January 2025, and the process of screening, inclusion, and analysis continued until mid-March 2025. To identify relevant sources, we used the Publish or Perish software to perform structured searches across two major databases: Google Scholar and Crossref.

Search queries were constructed using combinations of terms related to public attitudes and artificial intelligence, specifically:

- "public perception" AND ("artificial intelligence" OR "AI")
- "public opinions" AND ("artificial intelligence" OR "AI")
- "public attitudes" AND ("artificial intelligence" OR "AI")
- "public" AND "trust" AND ("artificial intelligence" OR "AI")

The combined search yielded over 1,000 items. These results were consolidated into a master spreadsheet and scored for relevance using a five-point system. One point was assigned for each of the following criteria:

- "AI" or "artificial intelligence" appeared in the title.
- "Public perception," "public opinions," or "public attitudes" appeared in the title.
- "AI" or "artificial intelligence" appeared in the abstract.
- "Perception," "opinions," or "attitudes" appeared in the abstract.
- Both "public" and "trust" appeared in the title.

Items were ranked in descending order of their relevance scores. Abstracts of highly ranked texts were then read more closely to identify and remove irrelevant results. Studies were excluded if they focused exclusively on technical features of AI, such as robotic perception (defined here as a robot's sensory capabilities) or system design methodologies for trustworthy AI without an empirical focus on public attitudes.

After this initial filtering, the shortlist of approximately 150 items was expanded using a snowball method: references cited within the reviewed articles were examined for additional relevant sources not captured by the keyword search. Newly identified items were assessed using the same criteria and, where appropriate, added to the dataset.

The final dataset includes 251 items published between 2011 and 2025, comprising peer-reviewed academic journal articles, conference papers, and grey literature such as public opinion surveys and industry reports. Each paper was manually coded across multiple dimensions within a structured dataset. The metadata captured included the authors, title, year of publication, outlet, abstract, and number of citations. We also recorded any explicitly stated theoretical or conceptual frameworks used by the authors, as well as the research questions, objectives, or hypotheses when directly formulated in the text. A concise summary of the study's main findings was provided under results. To capture the focus of each paper, we identified its general domain of application - such as healthcare, education, or security - and, where applicable, its specific use-case (e.g., facial recognition or military AI). Additional variables noted whether the study concentrated on the general public or expert populations, and whether it addressed national or public security issues. We also recorded the country or countries covered by the study, the method used (e.g., survey, survey experiment, focus group), and a brief methodological note describing the research design. Where available, we included the data collection period, sample size (N), and whether the study relied on secondary data. The dataset also captured the overall attitudes reported by respondents toward AI (positive, negative, or mixed), alongside other notable findings. Each study was assigned an overall quality score from 1 to 5, based on the outlet's standing, methodological robustness, sample size, and the study's geographic and thematic scope. Finally, we coded each paper's conceptualisation of the relationship between AI and society, using categories such as constructivism, instrumentalism, substantivism, critical theory, or technological determinism, depending on how the authors framed technological agency in relation to social structures and values.

## Domains and applications of AI

The organisation of the results section of this paper reflects the focus of the studies we reviewed. It is to say that the structure emerged inductively out of the review.

First, we start with the studies that were interested in surveying the perception of AI in general, i.e., without the focus on particular domains, use-cases, or applications. It does not mean that these studies do not refer to some sector- or type-specific reactions to AI – frequently they compare responses to specific uses of AI and show important similarities and differences in how AI is perceived in different contexts.

Secondly, we review studies of perception of AI in a specific domain. By domain of application, we mean a distinct sector of economy or of public life in which AI has been deployed. These sectors include education, healthcare, security, and public administration. In each of these sectors, AI has been used in different types of applications, such as automated decision making and autonomous vehicles and so also, we consider the effect of application as well as domain.

Thirdly, we review application-specific studies which focus either on AI-based solutions (such as autonomous vehicles) or on particular types of AI (such as generative AI).It means that a starting point for these studies in not the domain of application but rather particular solution.

While we acknowledge that this way of organising the results might not align, for example, with types of AI or machine learning models (such as the distinction between expert systems and neural networks-based AI or generative and discriminative machine learning models), we maintain that it is more feasible to assume a more phenomenological approach. It means that we not only present the landscape of perception of AI studies as it emerges from the reviewed literature but also that arguably the public perception of AI is formed at least partly through individual confrontation with AI application within different domains. In addition, we acknowledge the role of media and political narratives and frames in shaping the public attitudes toward technology.

## 3. Theoretical and analytical framework

### Constructivism

Our analysis is informed by constructivist approach to the relationship between societies and technologies. This approach, widely adopted in contemporary studies on science, technology, and society (STS), resists both instrumentalism – an idea that technologies are  mere tools – and technological determinism – an idea that society is shaped by technology – and instead traces the ways in which technology is developed and deployed in the manner that reflects moral, political, or cultural values. This is not to say that the technologies do not have certain features or characteristics that influence the way they are adopted and used but rather to claim that these features are to a large extent reflection of values embedded in the technology by its creators and these are the values that, in turn, might be perpetuated by technology and through technology shape the society (Vallor 2024).

Social constructivism is thus understood here not as an assumption that societies have complete freedom to shape technology, that freedom is always contained by established cultural, political, or moral norms which limit the space of possible uses of technology. Even though in principle it is always possible for society to decide, the technological impacts result from a combination of the essential features of technology, its design, situational context, and its use (Barney 2010). In addition, while we do not explore this aspect further in this paper, it must be noted that contemporary AI is being created in the context of a particular political economy. Technological corporations that offer AI-based tools to consumers, design these tools ostensibly to answer to some market needs or solve existing problems, although in the context of AI it has been claimed that it is a solution in search of a problem (Rosenberg 2024). In practice, the results of the review confirm existence of patterns in perception of AI across different publics and societies and dependence of those patterns on multiple factors that range from individual characteristics of users of technology and socio-cultural, contextual factors to technological properties of different AI solutions.

## Conceptual considerations

The conceptual landscape in the study of public perceptions of AI is marked by a proliferation of overlapping terms - perception, attitude, opinion, acceptance, and trust - each reflecting distinct dimensions of how individuals and publics relate to artificial intelligence. Although often used interchangeably, these terms denote conceptually different constructs. Perception refers to cognitive interpretations of AI's role and societal impact, while attitudes reflect evaluative orientations, ranging from optimism to scepticism. Opinions are more explicitly articulated judgments, typically shaped by prevailing media and political narratives. Acceptance denotes behavioural willingness to use AI, and trust emerges as a cross-cutting concept that not only influences but often mediates acceptance, particularly in contexts with high stakes or uncertainty. These distinctions are critical, as they signal that support for or resistance to AI is not merely a function of knowledge or technological literacy but is deeply embedded in cultural meanings, ethical considerations, and institutional relationships.

Figure 1 shows the mean uses - in the corpus of papers reviewed - of each of the main intentional verbs/terms. The striking observation is the dominance of "trust" which has a remarkable per paper usage over 100. From this alone we can assert that the concept public trust dominates the discourse in this field. Because only around 10% the papers in our dataset specifically focus on trust, this dominance would require further exploration. Specifically, we might need to ask, what are the reasons trust has become the organising idea in debates about the relationship between humans and AI?

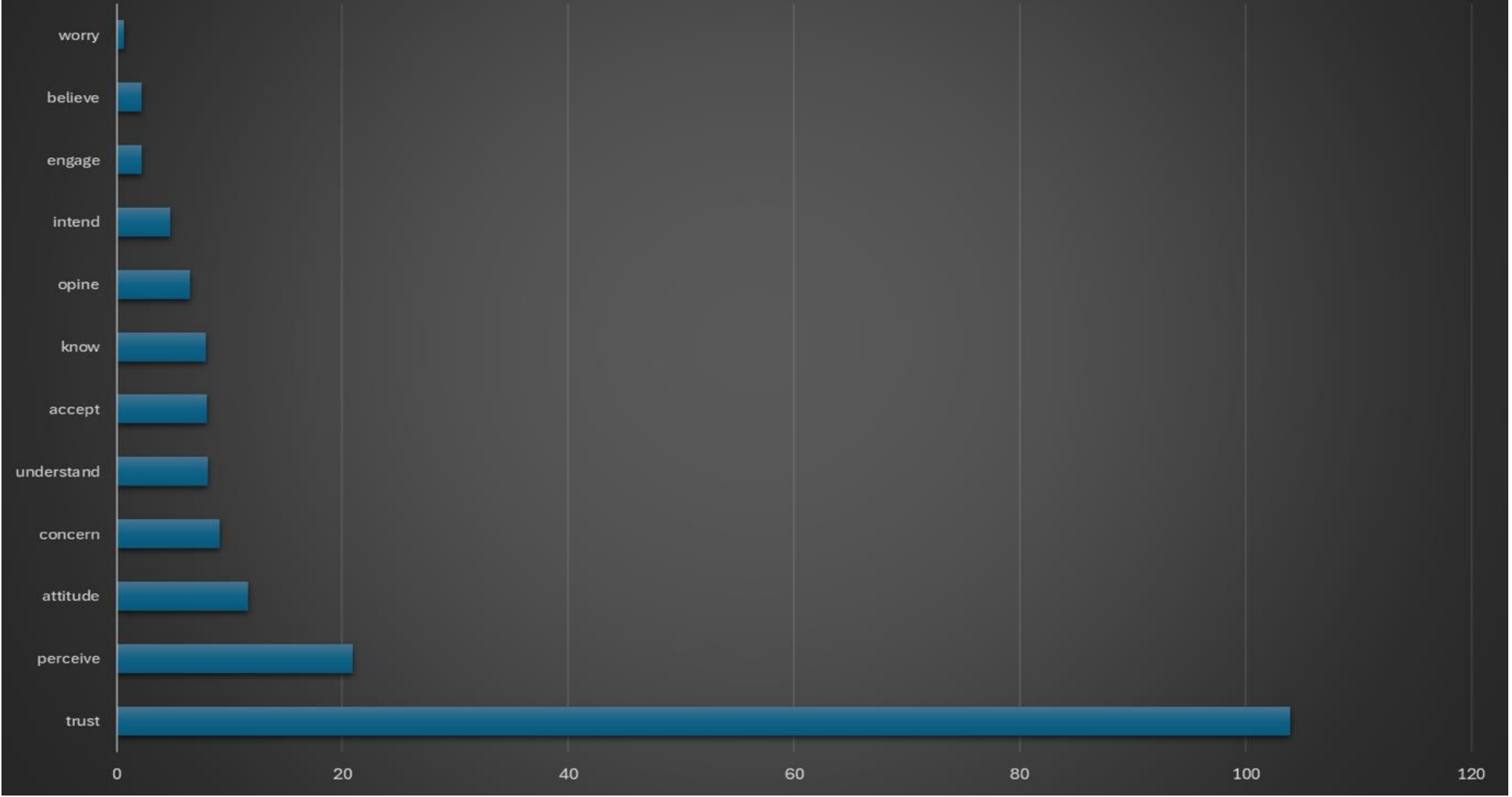

**Figure 1.** Mean uses per paper of the main intentional verbs/terms used to capture the publics relationship with AI.

It may be hypothesised that the dominance of "trust" in the reviewed literature underscores its centrality to debates about AI governance and deployment. Yet trust is itself a contested and multidimensional concept. In the AI context, it ranges from system reliability and technical robustness to interpersonal trust in institutions, developers, or even anthropomorphised systems like generative AI. Importantly, the reviewed studies reveal that trust is not reducible to technical features alone but is relational and contextual - it hinges on institutional histories, domain-specific imaginaries, and perceived alignment with public values. For instance, trust in AI for healthcare diagnostics may derive not only from perceived accuracy but also from trust in the healthcare system itself. This complexity highlights the need for conceptual precision and interdisciplinary dialogue in the field, particularly as researchers move from measuring "perceptions" to designing policies or systems intended to foster "trustworthy" AI. While in the reporting of the findings we follow the language of the reviewed sources, we acknowledge the need for clarification of these key analytical terms. Without such clarity, there is a risk of conflating normative aspirations with empirical findings, and of designing interventions that misinterpret the very public attitudes they intend to address.

## Factors influencing perceptions

In attempting to make sense of the diverse and often fragmented literature on public perceptions of AI we propose a structured analytical approach that distinguishes between three broad types of influencing factors: individual, contextual, and technical. This typology is not intended to impose rigid boundaries on inherently complex and overlapping influences, but rather it allows us to synthesise findings across different domains and disciplinary traditions.

**Individual factors** refer to demographic, cognitive, and value-based characteristics that shape how people perceive and evaluate AI. Numerous studies highlight the importance of age, gender, educational attainment, and socioeconomic status in explaining varying levels of AI familiarity, enthusiasm, or scepticism. In addition, deeply held ideological, religious, or moral beliefs can frame AI not simply as a technical innovation but as a socially and ethically charged phenomenon. These individual-level factors are often the most frequently studied in surveys, but their effects are not uniform - they are mediated by social context and the type of AI in question.

**Contextual factors** capture the broader social, political, and institutional environments in which individuals encounter AI. These include national and cultural settings, levels of democratic engagement, public trust in political and regulatory institutions, as well as the presence (or absence) of coherent national AI strategies and regulatory frameworks. Studies show that institutional trust can act as a proxy for confidence in AI systems themselves, and that perceptions of AI vary depending on whether it is used in contexts associated with care and fairness (e.g., healthcare, education) versus control and surveillance (e.g., policing, defence). Awareness of policies or public deliberations around AI also appears to influence how people evaluate the legitimacy of its deployment.

**Technical factors** refer to attributes and features specific to AI systems that shape how they are understood and judged. These include the transparency and explainability of AI decision-making processes, the degree of perceived autonomy or human-likeness in AI systems, and broader concerns about ethical risks, accountability, and labour market disruption. For instance, individuals may react differently to AI used in diagnostic tools than to AI embedded in social media algorithms, based on how intelligible, fair, or human-aligned these systems appear to be. Technical factors interact with both individual values and contextual trust frameworks, further reinforcing the need for an integrative analytical approach.

## 4. Results

### 4.1. Overview of the dataset

The literature review dataset consists of 251 items published between 2011 and 2025 (see Figure 2), with more 50% of papers published in 2023 or 2024, suggesting increased scholarly focus on studying perception of AI.

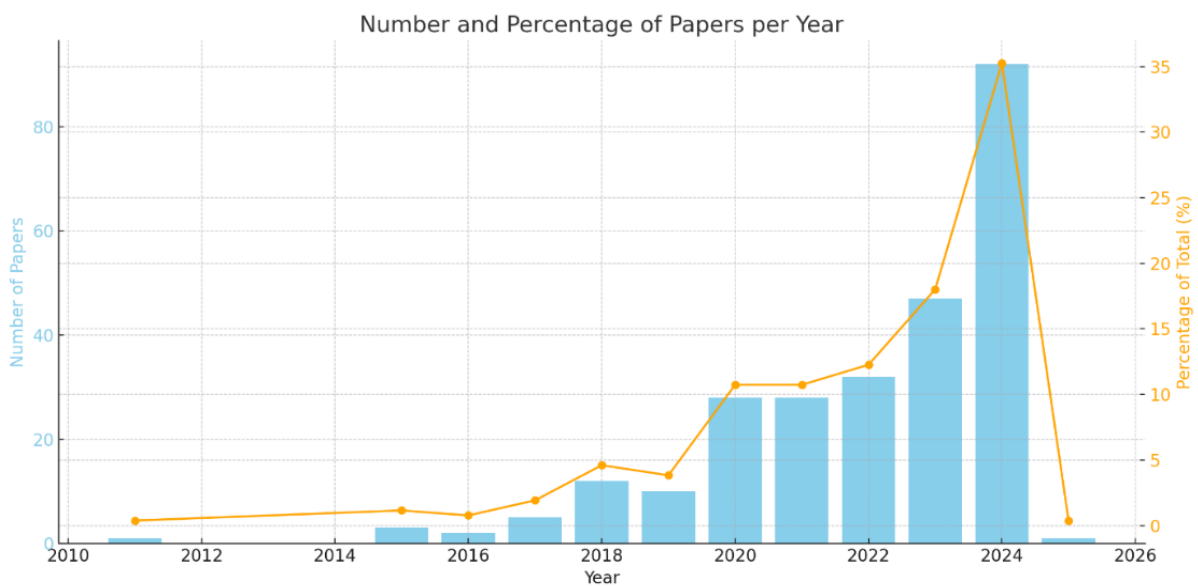

**Figure 2.** Publication year of the papers in the dataset.

The reviewed literature is heterogenous not only with respect to the methodologies and demographics it covers, but also in terms of topics that are the focus of research (see Figure 2).

A relative majority of the studies concern generic AI (not to be confused with artificial general intelligence or AGI a topic which has been – surprisingly perhaps - almost absent from the empirical studies) and concentrate on either the general public's perceptions of or trust in a nonspecific version of AI. The second group of papers studies attitudes towards AI and its particular use cases in specific domains, such as healthcare, education, public sector, security, business etc. The third group of papers focus on specific applications or types of AI, e.g., autonomous vehicles, automated decision-making, generative AI, and robotics. In addition, papers belonging to AI ethics and governance category focus either on theoretical analysis of ethical principles, norms, and values or on mechanisms of governing deployment or implementation of AI systems. It must be further noted that some of the papers fall into several categories. For example, if a cross-domain study about perceptions of AI in general asked questions about the use of AI in several domains, such as security or healthcare, it is also included in these categories.

In terms of geographic coverage, most studies focus on Western and Northern countries, followed by South and East Asia and Australia (see Figure 3). Specifically, 28.3% of papers present research on the United States, followed by Germany (9.6%), the United Kingdom (9.2%), China (8.4%), Australia and Japan (6.0% each), India (5.6%), and South Korea (5.6%). 10.4% of papers were comparative studies, covering multiple countries, and 13.5% were qualitative, theoretical, or literature review papers that did not offer empirical evidence from a specific country or region.

| Topic | Number of papers covering the topic | Percentage of N=251 |
|---|---|---|
| Generic AI perception and trust studies | 99 | 39.29% |
| Domain-specific studies | | |
| - Security | 29 | 11.5% |
| - Healthcare | 26 | 10.3% |
| - Public administration | 21 | 8.33% |
| - Education | 17 | 6.75% |
| - AI ethics and governance | 17 | 6.75% |
| - Business | 10 | 3.97% |
| - Art and design | 8 | 3.17% |
| Application-specific studies | | |
| - Automated decision-making | 18 | 7.14% |
| - Generative AI | 16 | 6.35% |
| - Robotics | 14 | 5.56% |
| - Autonomous vehicles | 10 | 3.97% |

**Table 1.** Topics covered by the reviewed papers

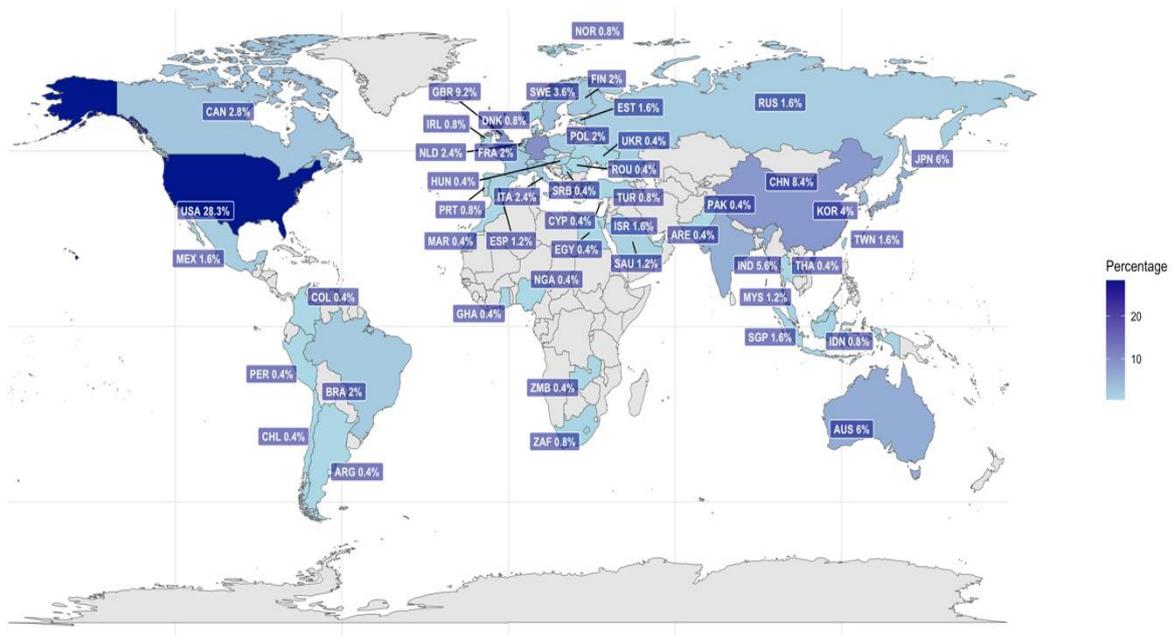

**Figure 3.** Percentage of papers related to each country.

In terms of research methods used in the studies (Figure 4), the majority (nearly 60%) relied on surveys (either online, face-to-face, or telephone). 12.4% of papers used survey experiments, followed by interviews (8.4%), qualitative methods (6.0%), social media data analysis (4.4%), and news media analysis (3.2%). A significant portion of the papers were theoretical studies (8.4%) and literature reviews (7.2%).

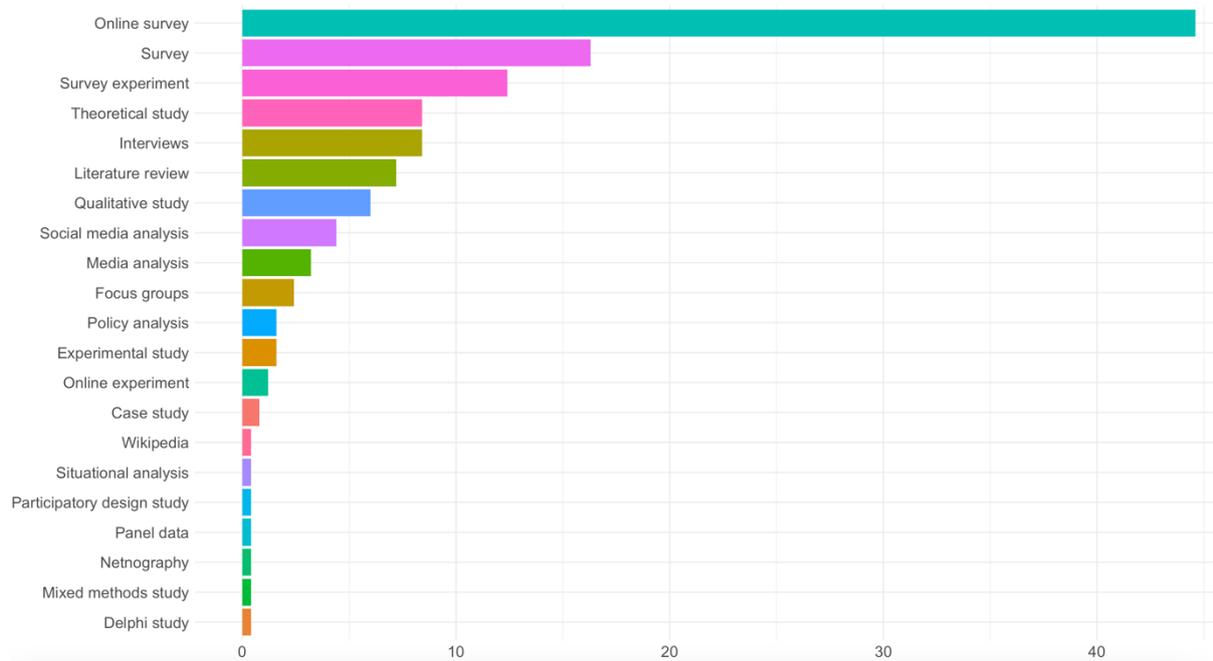

**Figure 4.** Distribution of research methods used across reviewed studies (in %).

## 4.2. General perceptions of AI

The largest number (99 or 39.3%) of the reviewed papers study perception of AI in general, without focusing on types of AI or application domains. The majority of studies in this category focus either on perceptions or on trust in AI by the public. A significant number of papers report on how media represent AI and what narratives drive perceptions of that technology.

### Perception of AI studies

Papers in this category focused mainly on perceived benefits and concerns regarding AI in general. This category included 29 studies published between 2016 and 2025. Geographically, the largest number of these papers covered the United States, European countries and Australia, with several multi-country or global studies and a significant portion focused on China, Japan, South Korea or India.

The review of the studies reveals several emerging patterns regarding positive and negative perceptions of AI applications and in terms of factors influencing support or concerns. It also shows that perceptions vary by geographic location and domain of application.

*Perceived benefits*
Perceived benefits of AI are most frequently associated with improved efficiency, enhanced public services, and practical utility in specific domains.

Across several studies, healthcare consistently emerges as the domain where respondents report the highest expectations for positive impact. In a 2016 poll (60 Minutes/Vanity Fair poll 2016), 44% of Americans selected healthcare as the field most likely to benefit from AI. Similarly (Kelley et al. 2021) found that 56% of respondents across eight countries viewed AI in healthcare as beneficial, a sentiment echoed in global studies (Ajitha, Geevarathna, and Huxley 2024; Arm Ltd 2020). In developing countries such as India and Nigeria, AI is often associated with modernisation and progress, particularly in healthcare and education (Kelley et al. 2021).

Efficiency and convenience also constitute central themes in positive perceptions. In the UK, respondents indicated they would use AI to save time (44.6%) and reduce errors (29.3%) (Holder, Khurana, and Watts 2018). Survey participants in Australia described AI as time-saving and innovative, particularly in the context of digitalisation and customer service (Regona et al. 2022). AI was also positively evaluated for automating repetitive tasks and simplifying daily routines in studies conducted in the United States, South Korea, and Japan (Arai 2018; Cave, Coughlan, and Dihal 2019; Kelley et al. 2021).

Other benefits include the potential for AI to support public services and infrastructure. In Japan, respondents supported the use of AI in disaster prevention and military operations, provided human moral involvement remained where socially valued (Arai 2018). In Australia, AI was viewed as useful in urban planning, emergency response, and infrastructure monitoring (Yigitcanlar, Degirmenci, and Inkinen 2024). Some studies also highlight favourable attitudes toward AI that augments human capabilities, such as predictive home maintenance systems or assistive technologies (Arm Ltd 2020).

*Perceived concerns*
Concerns about AI focus primarily on job displacement, privacy violations, and loss of control, with ethical unease emerging in specific contexts.

Fear of job loss remains a persistent issue. In the UK, 29% of respondents stated they would avoid AI because they feared it might replace their employment (Holder, Khurana, and Watts 2018). Globally, 53% of respondents believed that AI would result in more jobs being lost than created (Kelley et al. 2021). Similar sentiments were documented in Spain and Australia (Albarrán Lozano, Molina, and Gijón 2021; Regona et al. 2022), where perceived threats to employment correlated with more negative attitudes.

Privacy and surveillance are another major source of concern. In a 2020 global survey respondents expressed discomfort with AI systems perceived as autonomous or intrusive, such as smart locks or emotion-detecting devices (Arm Ltd 2020). In the same study, data security in cloud-based AI applications was a particular concern. A 2023 study (Park and Jones-Jang 2023) found that security and surveillance risks significantly reduced trust in AI among U.S. respondents, especially in healthcare contexts. Similar concerns were recorded in Germany (Ikkatai et al. 2024), where AI applications in facial recognition and surveillance were met with scepticism.

Some respondents are also uneasy about the ethical consequences of AI decision-making. Japanese respondents opposed the use of AI in child-rearing and interpersonal care (Arai 2018), emphasising the moral importance of human roles. In the U.S., respondents raised concerns about AI systems failing in critical infrastructures or being used by untrusted companies, such as Facebook (B. Zhang and Dafoe 2019). Existential risks, while less widely shared, were still notable: 34% of Americans believed that high-level AI might be harmful, with 12% fearing catastrophic outcomes (B. Zhang and Dafoe 2019).

*Factors influencing perception*

Perceptions of AI are shaped by demographic characteristics, direct experience, and national or cultural context.

Age, gender, and education level all influence public attitudes. Younger individuals are generally more accepting of AI, as found in the UK (Holder, Khurana, and Watts 2018), Australia (Selwyn and Gallo Cordoba 2022), and Serbia (Budic 2022). Higher education correlates with more favourable perceptions in Turkey and Spain (Albarrán Lozano, Molina, and Gijón 2021; Toksoy Çağal and KeskiN 2023), often due to greater understanding and technical confidence. While some studies find negligible gender differences (Gillespie et al. 2023), others report that men are more optimistic (Sartori and Bocca 2023), while women express more ethical and safety concerns (Park and Jones-Jang 2023).

Experience with AI technologies also moderates perception. In Australia, participants exposed to AI were more willing to reassess their views (Selwyn and Gallo Cordoba 2022). Another study (Ajitha, Geevarathna, and Huxley 2024) found that knowledge of AI's functionality increased satisfaction and support. A study in the United States showed that those who perceived ChatGPT as more sentient had more favourable attitudes toward AI, underscoring the role of perceived capability and anthropomorphism (Gupta 2023).

Geographic variation is also pronounced. According to a comparative study (Rice 2025), respondents in India (84.5%), Singapore (82.7%), and Taiwan (80.1%) demonstrated the highest support for AI, often linked to institutional trust and optimism about technology. In contrast, support was lowest in France (43.4%), Czech Republic (53.0%), and Poland (53.9%), where ethical concerns, economic insecurity, and distrust of institutions were more prominent (Rice 2025; Wilczek, Thäsler-Kordonouri, and Eder 2024).

Cultural predispositions, particularly uncertainty avoidance, also play a role. Individuals in countries with higher uncertainty avoidance, such as Poland, France, and Italy, are more likely to demand strong regulatory oversight (Wilczek, Thäsler-Kordonouri, and Eder 2024). Political ideology further influences regulation preferences, with left-leaning individuals favouring government oversight and right-leaning respondents more supportive of industry self-regulation.

## Trust in AI

Papers in this subcategory included 29 studies published between 2017 and 2025 and focused mainly on trust – defined in several diverse ways – in AI and factors that might increase or decrease it. Below we present, first, theoretical background from selection of the studies, and then factors found to be either increasing or decreasing trust in AI.

The theoretical studies offer conceptual and normative reflections on the meaning of public trust in AI, underlining that trust extends beyond technical performance to include political, relational, and institutional dimensions. One contribution (Knowles and Richards 2021) argues that trust in AI should not be understood solely in terms of system reliability but must be treated as a political relationship between institutions and publics, requiring transparency, accountability, and deliberative engagement. Another study

(Benk et al. 2024) identifies three prevailing approaches in the literature: trust as a design feature, trust as a matter of governance, and trust as a relational construct, and calls for clearer normative distinctions between promoting trust and justifying it.

Other contributions address how media and cultural narratives shape theoretical understandings of AI trust. One study (Ho and Cheung 2024) revisits classical communication models to argue that trust in AI is not a static belief but a mediated interpretive process shaped by media representations and public discourse. Another study (Nah et al. 2024) analyses global patterns in AI-related news consumption, suggesting that trust is shaped by transnational flows of information and culturally specific understandings of technological authority. Overall, these studies position trust as a socially constructed and contested concept, embedded in broader sociotechnical imaginaries and institutional frameworks.

### Factors increasing trust

Empirical research identifies several psychological, social, and technical conditions under which public trust in AI increases. A consistent finding across studies is that exposure to and familiarity with AI systems correlate with increased levels of trust. This pattern was evident in the United States (Ajitha, Geevarathna, and Huxley 2024), where respondents who had prior experience with AI expressed greater willingness to accept its use in public domains. Similar associations were documented in India where awareness of AI's functionality was linked to both higher satisfaction and trust.

Trust is also supported by explainability and transparency. In the United Arab Emirates, (Shin 2021) explainability emerged as a significant predictor of trust in AI across both public and student samples. This finding is showed in a broader empirical review (Glikson and Woolley 2020), which identified system tangibility, reliability, and transparency as essential to establishing trust in AI technologies.

Positive institutional environments also appear to foster trust. In Taiwan (Chen and Wen 2021), public confidence in both government and corporations was associated with greater acceptance of AI. A large-scale global comparison (Rice 2025) similarly found that institutional trust, support for scientific funding, and frequent engagement with science news were reliable predictors of trust in AI. In Nordic countries, AI policy frameworks explicitly prioritise public trust through communication and transparency, though the extent to which this improves perceptions is context-dependent (Robinson 2020).

Individual characteristics, including age and psychological traits, further influence trust. In the U.S., secure attachment style (Gillath et al. 2021) and higher levels of interpersonal trust (Choung, David, and Ross 2023) were both positively associated with AI trust, suggesting that trust in technology may reflect broader social trust dispositions.

Design features and ethical alignment play a central role in shaping trust. A meta-analysis of 65 empirical studies of AI (Kaplan et al. 2023) concluded that perceived fairness, user control, and ethical design are central factors contributing to higher levels of public trust.. Other research has emphasised that perceived usefulness and cultural alignment also influence trust levels (Yang and Wibowo 2022).

### Factors decreasing trust

Trust in AI tends to decrease when systems are experienced as opaque, unaccountable, or inconsistent with users' social expectations and values. Distrust is especially acute in collaborative scenarios where AI is expected to replace or act alongside human decision-makers. In a US-based study (Scharowski et al. 2023), trust in human–AI teams was found to be substantially lower than in human–human teams, even when AI performance matched human performance.

Low explainability and system opacity consistently undermine trust. When AI outputs are not readily interpretable or verifiable by users, scepticism increases. In the UAE, lack of explainability significantly reduced trust, especially in applications with high social stakes (Shin 2021). A multi-country review (Glikson and Woolley 2020) also found that insufficient transparency and perceived randomness in AI decision-making reduced trust.

Political and institutional distrust remains a barrier to trust in AI. In Taiwan (Chen and Wen 2021), low trust in government and commercial institutions corresponded with lower trust in AI technologies. In Germany (Luca Liehner et al. 2023), citizens were wary of AI systems perceived as being controlled by unaccountable actors, even when technological benefits were acknowledged. Broader cross-national studies (Rice 2025) show that institutional confidence is a mediating factor: when trust in national institutions is weak, people tend to extend that scepticism to AI systems as well.

Media exposure and regulatory uncertainty may also contribute to mistrust. While exposure to science news can raise familiarity and improve perceptions (Rice 2025), studies also suggest that negative or exaggerated portrayals of AI in popular media can heighten concerns about control, autonomy, and ethical risk (Yang and Wibowo 2022). Nordic countries have designed AI governance frameworks to explicitly mitigate these concerns, although their effectiveness remains under study (Robinson 2020).

### 4.3. Domain-specific results

This section presents an overview of domain-specific findings from the reviewed literature, highlighting how perceptions of AI are mediated by sector- or application-related factors such as ethical sensitivity, perceived risks and benefits, prior experience with technology, and the degree of human contact expected in each context.

## Healthcare

Perceptions and attitudes towards AI in healthcare domain have been studied in the context of AI in healthcare systems in general and of specific use cases, such as: diagnostics, robotic surgery, medical data handling and AI-based decision-making systems. In what follows we look at studies of each use case, starting with patients' or public's perceptions and then at perception of healthcare professionals.

*General public perceptions of AI in healthcare*
Studies on public perception of AI in healthcare tend to concentrate on factors influencing trust towards particular AI applications – including technical factors, such as human-like characteristics of AI – as well as perceived benefits and concerns. The studies reveal important cultural differences in acceptance of AI in this domain. A nationwide 2023 survey of 2,000 Japanese citizens (ages 20–69) (Mantello et al. 2024) found that while there is general trust in the healthcare system, AI and emotionalised AI (a type of AI that involves computer vision, deep learning algorithms, big data, natural word processing, voice tone analytics and biometric sensors and actuators) literacy remain low, with respondents wanting more government and media-provided information. Acceptance of AI in healthcare is influenced by gender (men are more positive), income and education (higher levels correlate with greater AI openness), and age (younger respondents, especially women, are less fearful). Despite recognising AI's potential benefits, many prefer human caregivers and are concerned about privacy, costs, and AI's impact on doctor-patient communication. These findings stressed the need for targeted AI literacy efforts and culturally sensitive AI integration to address scepticism and enhance acceptance. The national and cultural context of this study is crucial because the Japan's aging population, labour shortages in healthcare, and cultural emphasis on empathy in medical care create a unique demand for AI technologies that can simulate human-like emotional interactions while addressing public concerns about trust, privacy, and accessibility. A 2022 Indian online survey (Dash 2022) showed that cognitive trust is a stronger predictor of willingness to adopt AI-powered healthcare systems than affective trust. Factors such as expertise, product performance, and reputation significantly influence cognitive trust, while integrity and reputation influence affective trust. Benevolence does not significantly impact trust. Public trust then could be considered essential for adoption, with the results emphasising the role of transparency, performance reliability, and ethical considerations.

Several studies specifically concentrate on the use of AI in diagnosis. A 2020 survey-based US study (with a limited, convenience sample of 264 state fair participants) on the perceptions of AI and robotics in diagnostics and surgery found that respondents had nearly equal trust in AI vs physician diagnoses while 55% were uncomfortable with automated robotic surgery (Stai et al. 2020). The trust in AI's over doctor's cancer diagnosis was significantly higher. The study suggested participants opinions and confidence in AI are very sensitive to how AI is described – the confidence in diagnosis was greater in those respondents who received an indication that AI can make a diagnosis. At the same time, 88% of participants believed wrongly, that surgical robots currently operate autonomously in standard practice, thus suggesting existence of misperception of actual capabilities of AI technologies. Another 2020 online survey-based US study on trust towards AI vs doctor diagnoses (Juravle et al. 2020) showed different results, with participants trusting human doctors more than AI doctors, with the trust gap larger for high risk diseases. Trust in AI was particularly low when AI served as primary diagnosis tool. People were less likely to accept an AI-provided second opinion, especially for serious medical conditions. Respondents did not hold AI doctors to higher performance standards than human doctors. The study further found that informing participants about AI's superiority in diagnosis did not boost trust in AI, though it was "dramatically increased" when people are offered a choice between AI and human doctor, while heavily nudging them to opt for AI diagnosis, suggesting that giving participants a choice (or autonomy) make them "feel more in control" and "might alleviate(d) their hesitations about the AI (Juravle et al. 2020, 276). Another study on explainability of AI diagnostic systems found that providing local, case-specific explanations – which included rationales, visualised probabilities and real-world examples – increased trust and satisfaction compared to AI systems that did not provide such explanations or only offered text-based justifications (Alam and Mueller 2021). Global explanations about AI diagnostic process in general did not significantly improve immediate trust but enhanced long-term understanding of AI decisions. These results are further confirmed by a 2022 British qualitative participatory study (Banerjee et al. 2022) that looked at patient and public involvement as a method of building trust in AI. The authors suggested that engaging patients with lived experience of diseases and creating Research Advisory Groups that would enable patients to co-design algorithms and learn about the predictive models used in diagnostics and decision-making with the use of visualisation tools. Importantly though, the authors note that patients participating in the study – conducted at the Cambridge University Hospital – "may have had implicit faith in the institutions" and the same can be "very difficult to achieve in places where there is an existing trust deficit in doctors and hospitals, especially in low- and medium-income countries" (Banerjee et al. 2022, 3). A 2022 study with two online surveys on representative samples of German population (Wittal et al. 2022) revealed generally positive but cautious attitude towards AI in healthcare. In particular, 72% respondents recognized potential benefits of AI in diagnosis and therapy but expressed concerns about data privacy and ethical implications. Higher level of education correlated with greater familiarity and positive attitude towards AI. In line with other research, majority of respondents were more willing to pass on their data to public institutions than to private companies.

Another subset of studies focuses on the use of AI to support medical decision-making. A qualitative, interview-based 2023 British study of trust towards the AI-based decision-making systems in the context of intrapartum care (Dlugatch, Georgieva, and Kerasidou 2023) concluded that public institutions (such as universities and NHS) were more trustworthy as AI developers than private companies, emphasising priority of public good over profit. Participants of the study stressed the need of "trustworthy data" (reliable, unbiased, and inclusive, especially of BAME population) to perceive AI system that the institution developed as trustworthy. Furthermore, the participants agreed that AI's role should be supportive and that "people should have the final say in any decision-making process" (Dlugatch, Georgieva, and Kerasidou 2023, 10). The authors of the study concluded that the "trustworthiness and ultimate success of AI is not dependent on its technology alone but on people - the user groups who are positioned to delineate the ethical values to preserve and maximize, and the developers, policy makers, and healthcare professionals who translate these values into design and practice" (Dlugatch, Georgieva, and Kerasidou 2023, 13). A 2023 online survey in Britain (with a small sample of 172 laypeople) on using AI in deciding on allocating livers for transplant (Drezga-Kleiminger et al. 2023) showed 69.2% of respondents accepting such use with almost 72.7% saying they would not be less likely to donate their organs if AI was used In allocation. Participants valued accuracy, impartiality, consistency, interpretability, and empathy as important characteristics of decision-makers. Almost 62% of participants agreed that AI would lead to dehumanisation of healthcare, which proved to be the strongest predictor of acceptance of AI in this context (meaning that the more people thought AI leads to dehumanisation of healthcare, the more unacceptable they found using AI for liver allocation). Other impactful predictors were perception of AI as more or less likely to consider individual nuances of individual situations and the extent to which AI was perceived as more or less biased than humans.

*Expert perceptions of AI in healthcare*

Several studies on perceptions of AI in healthcare focus on domain experts, including doctors, nursing staff, and medical students. A 2021 online survey of perception of AI and robotics among medical students and faculty at the University of Nicosia (Cyprus) (Sassis et al. 2021) showed that that more than 75% of each group was in favour of including AI and robotics in the curriculum. Almost 60% of students and 48% of faculty agreed that physician's opinion should be followed in the case of difference between AI's and doctor's judgement, and dehumanisation of medicine was viewed as the most significant drawback of AI and robotics in the healthcare, according to almost 55% of students and 48% of faculty. Familiarity with the technology made students more enthusiastic about working in their medical speciality. A 2023 survey among nurses in Alexandria (Egypt) showed moderate attitudes towards AI in this group, while ethical awareness was a significant predictor of innovative work behaviour but did not have a very strong impact on attitudes towards AI (Atalla, El-Ashry, and Mohamed Sobhi Mohamed 2024). It suggests that the factors shaping attitudes towards AI, such as direct experience, training or perceived usefulness might be more impactful than ethical awareness alone. A 2019 review of studies on radiologists' perceptions of AI found a broad spectrum of views, ranging from enthusiasm to scepticism, anxiety, and fear of job losses (Pakdemirli 2019).A 2023 interview-based study of Swedish, Finnish, and German opinion leaders (politicians, insurance companies, media) on trust in care robots (Hoppe et al. 2023) identified four trust categories: trust in healthcare systems, trust in regulations, trust in technology, and interpersonal trust. Institutional trust is weakened by fragmented responsibilities, insufficient regulation, and a lack of structured information and implementation processes, while progressive trust in technology is low due to limited user experience, negative public attitudes, and a mismatch between design and real care needs. Dispositional trust varies by individual, with openness to care robots influenced by personal care needs, prior tech experience, and resistance to changing familiar routines.

The review of the studies in the healthcare reveals that despite geographical or cultural differences, some patterns can be observed among both the public and domain experts. In particular, the human factor – relating either to features of AI applications or to the role of human judgment in diagnostics or decision-making – seems to be important across different groups and populations. Similarly to other domains, perception of AI fairness as well as familiarity with the technology are important factors influencing perception and trust.

## Public administration

Studies reviewed in this section focus on three broad areas that are being impacted by introduction of AI: 1) trust, perception, and legitimacy of using AI by public institutions; 2) public services delivery; and 3) decision-making.

*Trust, perception, and legitimacy*

Studies on trust, perception, and legitimacy examine the conditions under which AI in public administration is viewed as fair, reliable, and institutionally acceptable. A 2020 global survey of respondents from 142 countries (Neudert, Knuutila, and Howard 2020) analysed perceptions of AI tools in decision making and found significant regional disparities in awareness and trust, with East-West divides in public attitudes. Concern about the harmful effects of AI is highest in North America (47%) and Latin America (49%), and much lower in Southeast Asia (25%) and East Asia (11%), with only 9% of respondents in China expecting that intelligent machines capable of decision making will mostly cause harm. Enthusiasm about AI decision making is strongest among business and government executives (47%) and other professionals (44%), while confidence is lower among manufacturing and service workers (both 35%).

A 2024 survey study in Portugal (Martinho 2024) identified transparency and strong regulation as the most important elements in building trust in AI used in public administration and minimising perceived risk. Perceived usefulness and behavioural control over AI were significant predictors of trust. Another 2024 study in Thailand (Surapraseart 2024) found that trust in AI in government was largely dependent on overall trust in government and was associated with perceptions of transparency, accuracy, and reliability. Similar results appeared in a 2024 study in Taiwan (Y.-F. Wang et al. 2024), which found that the "trust transfer" occurs between administrative processes, local government, political leaders and AI-enabled governmental systems. A 2020 experimental study in Germany (Starke and Lünich 2020) compared citizen perceptions of AI and human decision-makers in public administration and found that AI was seen as less legitimate overall, though legitimacy increased when AI supported rather than replaced human decision-making; fully autonomous AI-based decision-making was consistently perceived as illegitimate.

Cautiousness or scepticism towards AI and desire for AI to be functional rather than human-like was visible in the results of a 2023 Finnish survey (Lehtiö et al. 2023) on how citizens perceive AI in the context of smart city – with scenarios such as personal AI assistants, smart traffic stops, feedback chatbots, sustainable office buildings and emotion tracking AI. The research identified five main themes in the responses: "I don't like them monitoring me", "I want maximum gain for minimum effort", "I don't want AI to mimic people", "I'll avoid using AI if I consider the risk too high", and "I don't need to be concerned about AI". A 2023 online experiment in Belgium (Kleizen et al. 2023) tested the effects of ethical AI information on citizen' trust in and policy support for governmental projects. It showed that general ethical AI information had little effect on citizens' trust, perceived trustworthiness, or policy support, with prior attitudes, such as privacy concerns and existing trust in government or AI, serving as stronger predictors, indicating that building trust in governmental AI projects requires long-term, comprehensive strategies that address deeper public concerns and experiences.

Another 2024 study in Germany (Aljuneidi et al. 2024) examined citizen perceptions of automated decision-making and found that more detailed explanations improved perceptions of fairness but did not influence willingness to reuse the system, while higher AI literacy was associated with greater adoption regardless of explanation levels. The study suggested that greater human involvement and appeal mechanisms could positively influence citizens' perceptions. A 2024 study of Italian civil servants (Giraldi, Rossi, and Rudawska 2024) revealed positive attitudes toward AI and generative AI, with respondents perceiving these technologies as augmentative rather than replacements for their roles. Gender, age, and job position significantly influenced perceptions, with younger employees and those in healthcare demonstrating higher willingness to adopt AI. The perception of AI's usefulness was a more significant predictor of adoption than ease of use.

*AI in public services*

Studies focusing on the application of AI in public services examine its perceived usefulness, limitations, and contextual acceptability in citizen-facing government functions. A 2020 experimental study in Japan (Aoki 2020) explored public trust in AI chatbots in administrative services and found that trust varied depending on the type of service, with higher trust in simpler, structured service contexts. For example, trust was significantly lower in parental support chatbots than in waste separation chatbots. A 2022 online survey comparing Australia and Hong Kong (Yigitcanlar et al. 2022) found that respondents from Australia were generally more optimistic about AI applications in urban services, while Hong Kong residents expressed more scepticism, particularly regarding AI transparency and government AI initiatives; support for AI in governance, security, and urban planning was higher among individuals who associate AI with analytics and prediction, while those who linked it to humanoid robots or dystopian scenarios tend to be more sceptical.

Responses to AI of domain experts – in this case, public administration managers or employees – were studied in several national contexts. A 2023 interview study with city managers in Australia and the United States (Yigitcanlar, Agdas, and Degirmenci 2023) assessed perceptions of municipal AI applications and found that while city managers recognised the potential for AI to enhance service delivery and engagement, implementation remained limited. AI adoption is constrained by a combination of technical, institutional, and social factors, including data bias, lack of regulation, limited funding and expertise, user resistance, legal uncertainty, and concerns about transparency, accountability, and the opaque nature of automated decisions. A two separate studies in Mexico (Millan Vargas and Sandoval-Almazán 2024; Ruvalcaba-Gomez and Cifuentes-Faura 2023) revealed that both civil servants and managers in public sector perceive AI as potentially beneficial but lack necessary skills and knowledge for its implementation, while also pointing out to absence of AI strategies within their institutions.

*AI supporting decision-making in public administration*
Research on automated decision-making (ADM) focuses on public and institutional attitudes toward the use of AI to make or support decisions in government settings. A 2024 study in Estonia, Sweden, and Germany (Kaun, Larsson, and Masso 2024) revealed national differences in awareness, trust, and perceived suitability of ADM in public administration, which mapped onto historical differences in welfare regimes; the highest trust and awareness of ADM were found in Germany, followed by Estonia and Sweden, while risk perception highest in Sweden, moderate in Estonia, and lowest in Germany. A 2024 interview-based study in Sweden (Berman, De Fine Licht, and Carlsson 2024) focused on public sector professionals in the context of assessment of job seekers' need for support and found mostly negative reactions, with attitudes towards AI undermined by lack of transparency, interpretability, AI literacy, and stakeholder engagement. The study found that economic pressures for efficiency often override ethical considerations, contributing to decisions that do not always align with jobseekers' best interests. These findings are confirmed by a 2021 interview study in Finland (Drobotowicz, Kauppinen, and Kujala 2021) which examined public perceptions of trustworthy AI services in public administration and found that transparency and explainability were the most frequently mentioned conditions for trust, while opaque systems were associated with risk and reduced acceptance; respondents required control of personal data usage and involvement of humans in AI services.

The reviewed studies suggest important global differences in perceptions of AI in public administration, while also pointing out to some patterns, such as the requirement that AI applications and their decisions will be transparent, understandable, limited to routine tasks, and that they could not replace human accountability and presence; more broadly, several studies suggest that trust in AI is conditioned upon the trust in institutions that deploy it.

## Education

The perception and attitudes towards AI in education have been studied in the context of several main use cases: AI in general, AI-based educational technologies and systems (including automated scoring tools), generative AI, and AI literacy. Most papers in this category relies on surveying teachers and students, with just a few studies on the public's perception of AI in education.

*Educators' perspectives*
Majority of the studies in the education domain focused on reactions from education professionals about prospects of introducing AI-powered educational technologies (AI-EdTech) into educational systems, their existing AI skills and knowledge as well as motivation to learn AI. A 2023 international survey of K-12 teachers (Viberg et al. 2024) across six countries (Brazil, Israel, Japan, Norway, Sweden, and the USA) examined factors shaping trust in AI-EdTech. The findings showed that teachers with higher AI self-efficacy and understanding perceived more benefits, fewer concerns, and reported greater trust in AI-EdTech. Geographic and cultural differences influenced trust, i.e., cultural dimensions – which vary across nations – such as uncertainty avoidance, long- vs. short-term orientation, masculinity vs. femininity, and collectivism vs. individualism, have significant impact on trust: higher trust in AI-EdTech is predicted by higher uncertainty avoidance, higher masculinity, and higher collectivism. In contrast, demographic factors such as age, gender, and education level had no significant effect. The study emphasized the importance of enhancing teachers' AI understanding and considering cultural values to support the adoption of AI-EdTech in schools. The importance of understanding is confirmed by a 2023 experimental study in the United States (Nazaretsky et al. 2022) that explored K-12 science teachers' trust in AI-EdTech. The results showed that teachers were more likely to trust and adopt AI-EdTech when provided with clear explanations of how AI systems make decisions and how these tools can complement rather than replace human educators. Addressing common misconceptions - such as fears of job loss and privacy violations - also reduced concerns. A targeted professional development program further improved teachers' understanding and willingness to integrate AI-EdTech into their classrooms. A 2024 mixed-method study of Russian teachers' perceptions of AI as educational tool (Sadykova and Kayumova 2024) showed that majority (91%) of educators see AI as interesting and useful tool, although sometimes complex, risky and not very smart. Most respondents reported low level of AI competences (with the relatively highest level among younger teachers) and infrequent use of AI, while expressing readiness for training. A 2023 survey of secondary school technology teachers in South Korea (Bang-hee Kim 2023) found high overall awareness of the necessity and feasibility of AI education, especially among high school teachers and those with moderate AI training experience. Teaching efficacy was higher among male teachers, high school teachers, and those with training experience, while it was lowest among teachers with over 20 years of experience. A positive correlation was observed between perception of AI education and teaching efficacy. A 2023-2024 survey study in China (C. Zhang et al. 2025) investigated how risk perception and trust influence pre-service teachers' acceptance of AI-based educational applications. The study found that while AI trust positively affected acceptance, perceived ease of use and usefulness were stronger predictors. Among rural pre-service teachers, perceived privacy and safety risks significantly reduced trust in AI, an effect which was not observed in their urban counterparts. The findings highlighted the importance of functional reliability, user-friendliness, and transparency for AI adoption, and emphasize the need to address urban–rural disparities in AI acceptance.

There is an emerging subset of research focusing on developing countries. A 2023 survey-based study conducted in Zambia (Subaveerapandiyan, Sunanthini, and Amees 2023) investigated the knowledge and perception of AI among library and information services professionals. The findings indicated that while participants held a generally favourable and encouraging attitude toward AI, they also expressed concerns about the potential for AI to replace librarians' roles and about the infrastructural and professional barriers to adopting AI technologies in Zambian libraries. A 2024 exploratory survey study in Nigeria (Sanusi, Ayanwale, and Tolorunleke 2024) examined the factors influencing pre-service teachers' intention to learn AI, aiming to inform the design of effective AI teacher education programmes. Basic knowledge of AI and subjective norms emerged as the strongest predictors of intention of learning AI.

*Students' perspectives*
Several studies focused on students' trust in AI in the educational context. A 2024 cross-sectional survey in Poland and the UK (Kozak and Fel 2024) concentrated on identifying sociodemographic factors predicting trust in AI among students. The results showed higher trust in AI among students in the UK, while male students showed more trust in AI than females, regardless of place of study. The students in higher years of study trusted AI more than less advanced students, suggesting that longer experience as students may translate into ability to see benefits of technological innovation in education. A 2020 study in China (Qin, Li, and Yan 2020) focused on identifying factors contributing to trust in AI-based educational systems (AIED). Based on the empirical results from interviews and netnography, it proposed a framework of three categories of factors: context-related (pertaining to educational environment), technology-related (characteristics of tools), and individual-related (personal attitudes and beliefs). It further pointed out that dependability – a technology-related factor – has a positive impact on users' trust in AIED systems, while users expressed doubts regarding such systems' ability to guide value for learners and cope with emergencies in one-to-many teaching, suggesting that teachers cannot overuse these systems in pedagogical practice. Interpretability was found to reduce users' perceived risks. Gender disparity in the use of AI-based tools was the goal of a 2023 survey study in Ghana (Ofosu-Ampong 2023), among higher education students. The findings showed that gender significantly influenced students' use of AI in education, with notable disparities in perceived innovation characteristics.

Some of the research focused how students react to particular types of AI applications in the educational context. A 2024 study in Germany (Lünich, Keller, and Marcinkowski 2024) focused on perception of Academic Performance Prediction (APP), which refers to

AI applications predicting students' performance based on students' data. In comparison to the general public, students were more likely to approve of using AI in different areas of their lives, but that difference disappeared when AI was suggested in the areas of social life that bring considerable risk to the individual. Compared to the German population, students show increased perceptions of the risks and damages that come with APP. Two other studies on students' use of generative AI, one in China, and one in the United States, revealed similar patterns. The first, a 2023 survey (Yican, Yuhua, and Beng 2024), examined Chinese undergraduates' perceptions and use of generative AI tools like ChatGPT. The findings showed broad support for using generative AI as an auxiliary educational tool, with students expressing optimism about its potential for personalized learning, essay writing, and programming support. However, participants also acknowledged that generative AI cannot replace traditional learning methods and raised ethical and technical concerns. The study recommends blended learning approaches, ethical usage guidelines, and ongoing AI education for both students and instructors to support responsible integration in higher education. The second, 2024 study (Tossell et al. 2024) involving undergraduate engineering students in the United States, examined their experiences using ChatGPT for an essay writing assignment. The findings revealed that while ChatGPT did not make writing easier, it transformed the learning experience, with students expressing mixed responses. Post-use, students reported greater appreciation of ChatGPT's educational value and increased comfort with its ethical use, however concerns remained about its accuracy and lack of feedback confidence. Students supported the idea of ChatGPT assisting in grading, but only under instructor supervision. Overall, perceptions shifted from viewing ChatGPT as a cheating tool to seeing it as a collaborative resource requiring human oversight.

The results of the reviewed studies suggest openness of both teachers and students to the use of AI in educational contexts, while suggesting existence of barriers at the level of institutions as well as educators' skills. Some results suggest that AI should be used in supportive capacity and cannot replace traditional teaching and learning methods, suggesting the importance of human factor in this domain.

## Security

Only a small portion of the reviewed papers (around 11.5%) specifically focus on perceptions or trust in AI in the security domain. However we also found security-related results in papers focused on AI in general or on other domains, in particular in research interested in public sector or automated decision-making, thus increasing the portion of relevant papers to almost 16% (43 items). The main AI use cases or contexts in the security domain were:

- **military AI**, including Autonomous Weapon Systems (AWS), Lethal Autonomous Weapon Systems (LAWS) and Remotely Operated Weapon Systems (ROWS),
- **surveillance** (including facial recognition technologies),
- **judiciary use of algorithms** (e.g., in crime risk assessment and judicial decision-making),
- disaster prevention and management.

*Military AI*

The use of AI in the military context has been studied over the last decade. Mid-2010s studies on the use of AI in military context found majority of respondents opposing such use. In 2015, a global online survey (Horowitz 2016) reported 67% of respondents indicating that all types of LAWS should be internationally banned, with 56% saying LAWS should not be developed or used, and 85% that they should not be used for offensive purposes. 71% of respondents would rather see their country use remotely operated (ROWS) instead of lethal autonomous (LAWS) weapons systems when waging war and 60% claimed they would rather be attacked by ROWS instead of LAWS (Open Roboethics Institute 2015). A 2015 online survey experiment study in the US reported found the opposition to autonomous weapon systems (AWS) was not absolute but highly dependent on contextual factors: if these systems were framed as protection for US soldiers and as more effective than existing alternatives, public support increased by to 61% in comparison to 41% support when AWS were presented as offensive and not more effective. Additionally, the support increased when AWS were presented as necessity and with the condition that other countries and/or violent non-state actors were developing such weapons. A British focus group study conducted in 2022 (Hadlington et al. 2024) identified four major themes in public perceptions of AI in defence: the role of humans in AI decision-making (concerns over bias, accountability and oversight), ethical concerns (AI acceptable in logistical support but not for lethal decision-making, trust in AI vs. trust in organisations (scepticism toward autonomous systems), and the role of media and misinformation (participants relying on media and social narratives about military AI). The study found that participants believed that AI adoption in military is already widespread. An online survey-based 2024 study on LAWS in Japan (Arai and Matsumoto 2024) revealed increasing acceptance of autonomous military technology. Participants emphasised the need to protect civilian casualties over military casualties but did not show significant preference for remote-controlled over fully autonomous systems. The latter finding suggests that human oversight may not be as crucial for public acceptance of military AI as previously considered. Respondents showed some preference for using LAWS in homeland defence, which indicates that national security concerns might justify deployment of such systems. Gender has been a significant predictor, with male respondents more likely to support LAWs, while age did not show correlation with support levels.

In the SPAAI survey 41% of respondent said that they saw AWS systems as being very or fairly beneficial with older males being the most likely to perceive benefits. However 71% said that they were very or somewhat concerned with that concern being evenly distributed across demographic groups, but with more educated respondents being more likely to express concern (80%) that those with No academic or vocational qualifications (64%). See appendix E for more details.

*AI in surveillance applications*
While military AI solutions were studied already in the mid-2010s, surveillance applications such as facial recognition technologies (FRT) became a research interest with the AI advancements in the last 5-7 years. An online survey four country study conducted in 2019 (Kostka, Steinacker, and Meckel 2021) found the highest acceptance of FRT in China (67%), followed by the US (50%) and the UK (48%), with German respondents showing the weakest acceptance (38%). Factors influencing the acceptance included prior factors (socio-demographics and experience with technology) and antecedent factors (perceived benefits, risks, usefulness and reliability). Prior factors proved less significant than antecedent factors: perceiving improved security as a consequence of FRT proved to be particularly strong, positive factor for explaining its acceptance across all countries, followed by improved efficiency and convenience, with Chinese respondents being more techno-optimistic in contrast with German public in which convenience was not associated with FRT acceptance. A 2020 representative survey in the US (Brewer et al. 2022) found that increased support for FRT applications by law enforcement was predicted by overall television viewing and crime media viewing, with Fox News viewing predicting support for FRT to monitor protests with little evidence that news media use as such being a predictor of attitudes toward FRT. A 2024 study on AI in different surveillance applications (FRT, risk-assessment algorithms, license plate readers, and mobile phone location tracking) in the US (Boudreaux et al. 2024) showed that security, accuracy, and privacy were the most important factors for the government uses of FRT, with convenience being the least important. Less than quarter of respondents trusted the government's use of FRT. Risk technology was supported by more than 80% of respondents in emergency assessment in cases such as hurricane, but less than 45% for assessing whether someone may commit a violent crime and only 36% to assess risk of someone violating immigration law. Among different uses of license plate readers (LPR), the tracking of movement of a vehicle suspected of being involved in a crime was supported by almost 80% in contrast to less than 30% of respondents supporting using it to identify owners of vehicles in voting locations. The use of mobile phone location tracking was supported by almost 70% of respondents when tracking someone who appears likely to commit a crime, with less than 50% of support in other cases, and higher opposition than support in tracking persons at protests, demonstrations, public events, and at voting locations.

*AI in judiciary*
Use of AI and algorithms in judiciary has been studied in the context of decision-making, crime or parole violation risk assessment, as well as automation of administrative tasks. A 2018 Pew Research Center study (Smith 2018) of the US public on algorithmic decision-making (ADM) found that majority of Americans are sceptical about the fairness and acceptability of ADM in key areas, such as finance, employment and criminal justice. Criminal risk assessment for people up for parole was acceptable for 42% with 56% founding it unacceptable. A more recent, 2024 study in the US (Kim and Peng 2024) concentrated on acceptance of AI-based judicial decision-making (AI-JDM) in comparison to human judges. It found a general aversion to AI-JDM with regard to perceived risk, permissibility, and social approval, while in cases rooted in the moral foundation of fairness, it received slightly higher social approval. Ethnicity and race affected the attitudes with racial and ethnic minorities showing more approval in AI-JDM than majority population. Overall, the results suggested preference for human judges over AI.

*Disaster prevention*
A 2015-2016 online survey in Japan on different AI applications found general support for machine involvement of AI in disaster prevention and military operations while rejecting automation in personal domains, such as child-rearing (Arai 2018). A 2021 online survey in Melbourne, Australia (Kankanamge, Yigitcanlar, and Goonetilleke 2021), found that younger generations (18–44) expressed a higher level of trust in AI for disaster preparedness, response, and recovery, whereas older adults (55+) exhibited scepticism, particularly regarding AI's role in assessing disaster damages and developing recovery plans. People with tertiary education were more likely to trust AI applications in disaster management, while those with lower levels of education remained cautious about AI's predictive capabilities. Public sector workers in administration, safety, and emergency services were more supportive of AI applications in disaster management, particularly in resource mobilization and early disaster prediction.

Papers reviewed in this section show that – as in other domains – individual, contextual, and technological factors significantly change perceptions of security-related use of AI. Military AI remains the most controversial, with some research suggesting the support for it slightly decreases, and remains lower in comparison to other types of AI in this domain (Horowitz et al. 2024).

## AI ethics and governance

Studies reviewed in this section focus on public attitudes toward AI ethics and governance. While the two areas are not identical, they overlap in the sense that they inform the way in which AI systems' development, deployment, and use should be managed by public and private organisations in compliance with ethical norms or legal regulations. The studies we reviewed are diverse in their focus: some of them investigated which ethical principles should be prioritised in the development of AI, while other studied public attitudes toward ethics and governance as such.

Studies on attitudes on AI ethics and governance focus on trust toward institutions that manage AI development, ethical frameworks for AI, and particular methods of ensuring ethical governance of the technology, such as algorithm auditing. A 2020 online survey study in the US (B. Zhang and Dafoe 2020) focused on Americans' perceptions of AI governance challenges and their trust in institutions to manage AI. It showed that respondents regarded all AI governance challenges presented to them as important, yet they expressed only low to moderate trust in governmental, corporate, and multistakeholder institutions. The US military and university researchers were the most trusted groups to develop AI, with the US government institutions ranking lower than tech companies. The US military and university researchers were the most trusted groups to develop AI, with the US government institutions ranking lower than tech companies. A 2024 online survey and interview study with international scope (Baldassarre et al. 2024) focused on AI practitioners' visions of Trustworthy AI (TAI) principles and the challenges they face in implementing them. It showed that despite the proliferation of frameworks and guidelines to support Trustworthy AI, practitioners encounter specific challenges and express needs for tools, knowledge, and further guidance. The study revealed that participants cared about privacy and transparency, with explainability and dataset quality improvements being the most used strategies to ensure trustworthiness, while business constraints being considered the impediment to develop trustworthy AI applications. Focusing on particular way in which trustworthiness of AI might be communicated, a 2023 interviews and survey study in Switzerland (Scharowski et al. 2023) focused on end-users' attitudes toward certification labels. It showed that such labels significantly increased end-users' trust and willingness to use AI in both low- and high-stakes scenarios - with stronger effects in high-stakes contexts. A 2024 interview study in the USA (Lassiter and Fleischmann 2024) focused on how AI auditing professionals approach the creation of calibrated trust in AI tools and audits. It showed that, given the diverse methods in the AI auditing ecosystem and prevailing information asymmetries, users' trust in AI systems can be diminished. The research identified key themes from interviewees' experiences and offered recommendations, such as auditor monitoring and effective communication, to rebuild trust in AI audits. As countries around the world take steps to regulate AI, researchers begin to look at specific policies. A 2024 policy analysis study in India (Biju and Gayathri 2024) focused on India's data policy drafts and AI-related policies, examining whether they reflect constitutional values. It showed that the policy drafts position AI as both a social problem solver and an economic growth incentive while warning of the potential social disruptions caused by algorithms. Furthermore, the analysis critiques these documents for being primarily crafted by industry stakeholders and technocrats, and it enquires whether they uphold constitutional values such as inclusion, diversity, rights, liberty, justice, and equality.

Several other papers focused on discourse about AI ethics in attempt to identify stakeholders who might have impact on ethical guidelines or policy regulations. A 2024 literature review study (Zhu 2024) looked at this discourse in Chinese academia. It showed that, in the short term, Chinese scholars' ethical concerns predominantly mirror international ethical guidelines, while long-term implications reveal significant cultural differences and a preference for strong-binding regulations over weak ethical guidelines. The review found that the academic discourse is dominated by male scholars from elite universities. Another 2023 study in China (Mao and Shi-Kupfer 2023) focused on analysing public discussions about AI ethics on social media platforms, showing that the discussions, drawn from platforms like WeChat and Zhihu, involved a diverse set of participants, including scholars, IT industry actors, journalists, and the general public, who addressed a broad range of concerns related to AI applications. The authors found that participants of online discussions frequently referenced foreign sources, indicating that international (mainly Western) deliberations were to some extent shaping Chinese discussions. Another 2024 survey study (Binst, Bircan, and Smets 2024) on the discourse about AI ethics in the European Union focused on the unequal communication of opinion regarding AI, a threat to survey validity and epistemic justice. It showed that higher social positions are more likely to communicate an opinion, with significant effects of self-perceived social class, political efficacy, and cultural capital. By applying Bourdieu's theoretical framework, the study suggested that habitus mediates this inequality and recommends that future surveys implement corrective measures to mitigate epistemic injustice.

Another subset of studies focused on attitudes towards particular ethical challenges and on understanding of AI ethics knowledge. A 2022 online survey study in Japan (Ikkatai et al. 2022) focused on public attitudes toward AI ethics across eight themes drawn from common AI guidelines: privacy, accountability, safety and security, transparency and explainability, fairness and non-discrimination, human control, professional responsibility, and promotion of human values. It showed that agreement or disagreement with AI applications varied by scenario, illustrated by notably high anxiety when AI was used with weaponry and that age was significantly associated with these attitudes, while the effects of gender and AI understanding differed by theme and scenario. The study showed that concern about each of the eight themes differed with regard to different scenarios, suggesting that ethical considerations depend to some extent on the context in which AI is being applied. A 2024 Wikipedia-based study with global scope (Wei et al. 2024) focused on how the public comprehends the body of knowledge on AI ethics through Wikipedia. It showed that, by applying a community detection approach, the study identified a hierarchical structure in which primary topics predominantly revolve around knowledge-based and ethical issues, for example, transitions from Information Theory to Internet Copyright Infringement.

While the reviewed studies significantly differ in terms of their goals and scope, it can be observed that in comparison to earlier AI ethics considerations in the academic literature, researchers started paying attention not only to the public's responses to AI ethics guidelines and governance mechanisms, but also to particular ways trustworthiness of AI is being established. As such, these studies could offer some actionable, practical insights into building trust in AI applications, in particular with regard to auditing methods.

### 4.4. Application-specific studies

#### Automated decision-making

Research on algorithmic or automated decision-making (ADM) focuses on perception and trust in AI-driven decisions, as well as the role and understanding of fairness of such decisions.

Literature reviews of studies related to ADM showed conceptual and methodological diversity in how the central notions of trust and fairness are understood. A 2021 methodological review of 83 empirical papers on trust in AI-assisted decision-making (Vereschak, Bailly, and Caramiaux 2021) found inconsistency in the way trust in AI had been studied. Trust definitions were often incomplete or even not provided. The three key elements of trust – vulnerability, positive expectations, and that trust is an attitude – were not always incorporated in the studies. Furthermore, the study revealed lack of standards in research design, measures, tasks, or procedures that have been used in assessing trust, with additional questions about the constraints of laboratory experiments for investigating the dynamics of trust. Another literature review of 58 empirical studies (Starke et al. 2022) observed that the data on citizens' perceptions of algorithmic fairness came almost exclusively from Western democracies. It also found heterogeneity in how algorithmic fairness was defined. In general, studies on the perception of fairness in ADM focused on four aspects: predictors of perception related to characteristics of algorithms, human predictors (related to sociodemographic characteristics of respondents), consequences of ADM, and on comparison between ADM and human decision-making. The key takeaway from that review as that perceived fairness of ADM system is highly context-dependent and determined by not only technical design of algorithms, but also by the domain of application, and by the specific task and its stakes. The authors noted that theoretical incoherence of the studies in this area contributed to inconclusiveness of the empirical results.

Regarding the context of application of ADM, a 2018 Pew Research Center survey in the United States (Smith 2018) found widespread scepticism toward ADM, with 58% of respondents feeling that computer programs feeling that computer programs will always reflect some level of human bias, while 40% thinking that it is possible to design the program in a bias-free way. In particular, context of decision and type of application matters: criminal risk assessment for parole was unacceptable to 56%, automated resume screening for job applicants to 57%, while automated video analysis of job interviews to 67% and personal finance score using many times of consumer data to 68%. Concerns raised by respondents included violation of privacy, unfairness, removal of human element and complexity of humans that cannot be captured by algorithmic systems. The importance of the type of decision that is made by ADM system was confirmed by a 2018 online experiment study in the US (Lee 2018). It found that perceptions of fairness and trust in algorithmic versus human decisions depend on task type: for mechanical tasks (e.g., work assignment or scheduling), both were judged similarly, though algorithms were associated with efficiency and objectivity, while humans were valued for their authority and potential for social recognition. In contrast, for human-centred tasks (e.g., hiring, work evaluation), algorithmic decisions were seen as less fair and more emotionally negative, due to their perceived lack of intuition and the dehumanising experience of machine evaluation.

Several studies focused on understanding the notion of fairness and of reasoning of people assessing fairness of ADM. A 2018 study in the US (Grgic-Hlaca et al. 2018) found that fairness concerns about using specific features in decision-making extend beyond discrimination and include considerations of relevance and reliability, though people often disagree on which features are unfair. These disagreements stem from differing assessments of the features' latent properties, especially their causal influence, but individuals tend to apply similar reasoning heuristics when forming fairness judgments. Similarly, another 2019 experimental study in the US (Srivastava, Heidari, and Krause 2019) attempted to identify the notion of fairness that best captures people's perception of fairness. It found that the most simplistic mathematical notion of fairness (demographic parity) most closely matched people's idea of fairness in crime risk prediction and skin cancer risk prediction scenarios in comparison to alternative, more complicated definitions of fairness. Furthermore, it showed that participants considered accuracy more important than equality in high-stake predictions. A 2020 study conducted in the Netherlands (Araujo et al. 2020) focused on the extent to which personal characteristics can be linked to perceptions of ADM. It showed that general knowledge (education) was positively associated with perceived usefulness of ADM, while domain-specific knowledge (e.g., programming, AI, algorithms) was positively associated with both usefulness and fairness, but neither form of knowledge predicted perceptions of risk. Online privacy concerns were linked to lower perceptions of usefulness and fairness and higher perceptions of risk, whereas online self-efficacy had the opposite effect. Age was negatively associated with usefulness and positively with risk, gender influenced usefulness (with females rating it lower), and belief in economic equality was positively associated with usefulness and fairness, but not with risk. Another 2020 study in the US (R. Wang, Harper, and Zhu 2020) investigated factors influencing perception of fairness, both on the part of algorithm (outcomes, development, and deployment procedures) as well as individual differences. It showed that people rate the algorithm as fairer when it predicted in their favour, even if it is described as being biased against particular demographic group. Education level, gender, and some aspects of development procedure moderated this effect: it was stronger for females and the smallest for participants with higher education levels; participants with lower computer literacy tend to perceive the algorithm as less fair.

Our review confirms conclusions from the two major literature reviews of perception of ADM studies cited above. First, we observe that perception and trust in the context of

ADM is highly contextual and relates to the domain of decision-making and the type of decision. Second, as in other domains, demographic characteristics matter, with younger, higher educated, male respondents perceiving ADM and its fairness more positively in comparison to other groups. Third, while there is no consistency in how the notion of fairness is defined, empirical studies show that understanding and assessment of fairness of ADM it is also context-dependent and related to social roles of respondents.

## Autonomous vehicles

Research on public attitudes towards autonomous vehicles (AVs) focuses on perceived risks and factors influencing acceptance of autonomous cars and autonomous transport systems, such as driverless trains.

Studies on autonomous cars report a range of individual, technical, and ideological factors that shape acceptance. A 2024 bibliometric review (Naiseh et al. 2024) identified three primary key research areas in the literature on AV trust and risk: behavioural aspects of AV interaction, uptake and acceptance, and modelling human-automation interaction. The analysis suggested that factors influencing people's perception and acceptance of AVs include not only features of such vehicles (e.g., predictability of their behaviour), but also previous experience and demographic characteristics. In line with these findings, an earlier, 2021 global literature review (Othman 2021) found that public acceptance of AVs hinges primarily on perceived safety, trust in technology, and ethical concerns and reliability of the technology, with general conclusion that people are being highly concerned about AVs and the level of fear increasing with the increase in the number of accidents reported. Previous experience with AVs' features influences positively public acceptance; in line with findings in other domains, gender and education levels predicting attitudes – males and people with high education being more positive toward AIs than females and people with lower level of education. These findings were confirmed another 2021 study in the US (Dennis, Paz, and Yigitcanlar 2021). which found that people with prior exposure to connected and autonomous vehicles were more likely to accept and adopt them. Young, highly educated, male respondents were, again, the demographic more positive toward AVs  Similarly, a 2022 survey in Saudi Arabia found younger and better educated respondents having higher expectations of benefiting from the transition to AVs (Aldakkhelallah et al. 2022).

Relationship between perceived risks and benefits of AVs tends to be complex. A 2018 study in the US (Koul and Eydgahi 2018) showed that perceived usefulness (PU) and perceived ease of use (PEOU) are both significantly related to behavioural intention to use AVs, with perceived ease of use being the stronger predictor. Gender, level of education, and household income did not have any moderating influence on the relationship between PEOU and PU, while with increase of age and years of driving experience, the intention to use driverless cars decreased. A 2024 study in Ireland (Cugurullo and Acheampong 2024) found that while individuals were largely afraid of AI-driven cars, they were nonetheless willing to adopt this technology as soon as possible, with male respondents being more positive that females. The authors of the study found that despite fear pushing people away from AVs, multiple perceived benefits – including individual, urban, and global social and environmental benefits – do the opposite, suggesting that the publics' acceptance of AI should be understood as resultant of diverse and sometimes contradicting factors.

Several studies suggest that attitudes toward AVs are correlated with other individual factors and roles, including ideologies and belonging to particular category of users. A 2021 study (Tennant 2021) – based on surveys conducted in the US, UK, and 10 EU countries – found that public attitudes towards AVs mostly consistent, with around 50% respondents feeling uncomfortable with the prospect of riding as a passenger or driving alongside AVs, while around 30% respondents declared being comfortable in these scenarios; reluctance to give up control and AV reliability were the most contributing factors. A 2020 study (Peng 2020) on the ideological divide in the perception of self-driving cars in the US found that political ideology significantly shapes public attitudes, with liberal individuals more likely to support AVs and conservatives more likely to express concern. Familiarity with AVs and scientific literacy reduced respondents' concerns and their support for restrictive regulation, but their effect was weaker among conservatives, which indicated that people may assimilate new information in a biased manner that promotes their worldviews. A 2017 Australian study (Fraszczyk and Mulley 2017) of attitudes toward projected driverless metro system in Sydney found that significant portion of respondents expressed concerns about safety and reliability, while showing diversity of attitudes depending on the preferred mode of transportation, suggesting that directed marketing and educational initiatives could be more efficient in informing different groups of the public about driverless trains. Similarly, a 2021 focus group study (Dogan, Barbier, and Peyrard 2021) with participants belonging to different categories of road users (drivers, pedestrians, road users with disabilities) expressed diverging opinions with regard to principles that should govern AVs. For example, pedestrians stressed need for equity by supporting the integration of AVs in public transport. The findings suggested, again, that inclusive policy approach is important to ensure public acceptance of the technology.

The studies in this domain suggest that attitudes toward autonomous vehicles are shaped by perceived usefulness, familiarity with the technology, age, education, and political orientation. As such, these findings are consistent with results obtained in other domains where AI is applied.

## Generative AI

Generative AI emerged as a result of advancements in foundational models, especially Large Language Models in the late 2010s. It became important subject of research in recent years, after tools such as OpenAI's ChatGPT – followed by similar solutions from other major tech companies – were made available to wider groups of users in late 2022. Generative AI poses important challenges that include questions about intellectual property and rights to art used in the training of the models as well as possible malicious use of deepfake videos in disinformation.

Across the studies reviewed, a pattern of cautious optimism emerges in public perceptions of generative AI technologies. Sentiment analyses of large-scale social media data (Koonchanok, Pan, and Jang 2024; Liu and Lyu 2024; Miyazaki et al. 2024) consistently show that attitudes toward tools like ChatGPT and AI image generators are generally neutral to positive, especially among users who engage casually or creatively with the technology. However, this positivity is often accompanied by specific concerns, such as job displacement, unethical use of artistic content, or privacy risks, that vary across occupational groups. Illustrators, cybersecurity professionals, and lawyers, for instance, expressed more negative sentiments, while educators and marketers were more positive. Moreover, sentiment trends are not static: initial enthusiasm for ChatGPT in China gradually gave way to increased concern, driven in part by media hype and reflections on social and ethical implications (Liu and Lyu 2024)

Media exposure and framing play a critical role in shaping these perceptions. One study (Brewer et al. 2024) demonstrated that consuming technology news or science fiction increases support for AI-generated art, while also heightening concerns about its societal impact. Similarly, experimental findings show that how AI is framed - whether through artists' concerns or as a source of outrage - can significantly influence public attitudes. This aligns with evidence from another study (Ahmed 2023) where users showed third-person bias in judging the influence of deepfakes, believing others are more susceptible than themselves, despite limited actual detection ability. These studies suggest that while public engagement with generative AI is widespread and often curious or creative, it is also shaped by deeper anxieties about trust, authenticity, and control, while concerns are mediated by professional identity, media narratives, and users' own perceived understanding of AI.

## Robotics

Studies in this section focus on three key themes: trust in human-robot interaction, shaped by robot performance, transparency, and task context; public and media perceptions, which evolve over time and vary by robot type and platform; and cultural framing, which influences how societies interpret and accept robots, with notable contrasts between Western and Eastern perspectives. These themes show that acceptance of robotics depends as much on social and cultural factors as on technical capabilities.

Across the studies reviewed, several consistent patterns emerge regarding public perceptions and trust in robots and AI systems, shaped by performance, transparency, cultural framing, and media representation. Trust is strongly influenced by the robot's performance and design features, while human and environmental factors have comparatively weaker effects. Quantitative meta-analysis (Hancock et al. 2011) reveals that robot-related (in contrast to human-related factors) characteristics are the most significant drivers of trust in human-robot interaction. However, trust is also shaped by context and task-specific elements. For instance, while robot errors reduce perceived reliability and trustworthiness, people's willingness to comply with robots' instructions depends more on the nature of the task, especially whether its effects are revocable or not (Salem et al. 2015).

Transparency and embodiment influence how people interact with data-collecting robots, particularly in privacy-sensitive scenarios. For example, embodied robots are more effective than kiosks at eliciting personal information (Vitale et al. 2018), but this advantage is moderated by how transparently the system communicates its data practices. Interestingly, transparent information improves user experience regardless of platform, though users express greater privacy concerns with kiosks than with humanoid robots providing the same disclosures. These findings suggest that embodiment may mitigate perceived privacy risks, especially when coupled with clear communication.

Public and media sentiment toward automation and robotics is largely positive, though nuanced by technology type and context. Analyses of large media corpora show that public attitudes have shifted from viewing robots as solely industrial machines to seeing them as multi-purpose, assistive, and social devices. Sex robots, however, provoke the most polarised responses, suggesting that emotional and moral dimensions strongly shape acceptance (Javaheri et al. 2020). A similar picture emerges in sentiment analysis of over 95,000 news articles on Robotic Process Automation (RPA) (Kregel, Koch, and Plattfaut 2021), where initial hype gave way to more realistic and stable assessments, indicating that RPA has entered a maturity phase without a dramatic backlash.

Cultural narratives also influence public attitudes in significant ways. In Italy (Operto 2019), attitudes toward robots are complex and sometimes contradictory, combining appreciation for technological advancement with concern about its societal implications. These perceptions are shaped by socio-demographic factors, economic context, and widespread myths about robotics, underscoring the need for inclusive dialogue and public education. In contrast, cultural representations in Japan position robots not as adversaries but as companions who share space and intention with humans. Visual media in Japan frequently depict robots and humans jointly attending to a "third item," signaling a relational framing of coexistence, unlike the more confrontational imagery found in Western cultures (Sakura 2022). The reviewed studies demonstrate that trust and acceptance of AI and robotics are contingent not only on technological attributes but also on broader social, cultural, and communicative contexts.

## 4.5. Common trends in perception

The review of literature on perceptions and trust in AI included studies reveals multiple consistent patterns in how individual, contextual, and technical factors might influence – in different proportions and mediated by potential benefits (both individual and public) and perceived concerns (including risks) and perceived concerns – perceptions and trust in AI (Figure 5). While these patterns are observed across multiple domains, there are also some domain-specific caveats that might inform development and deployment of AI in specific contexts.

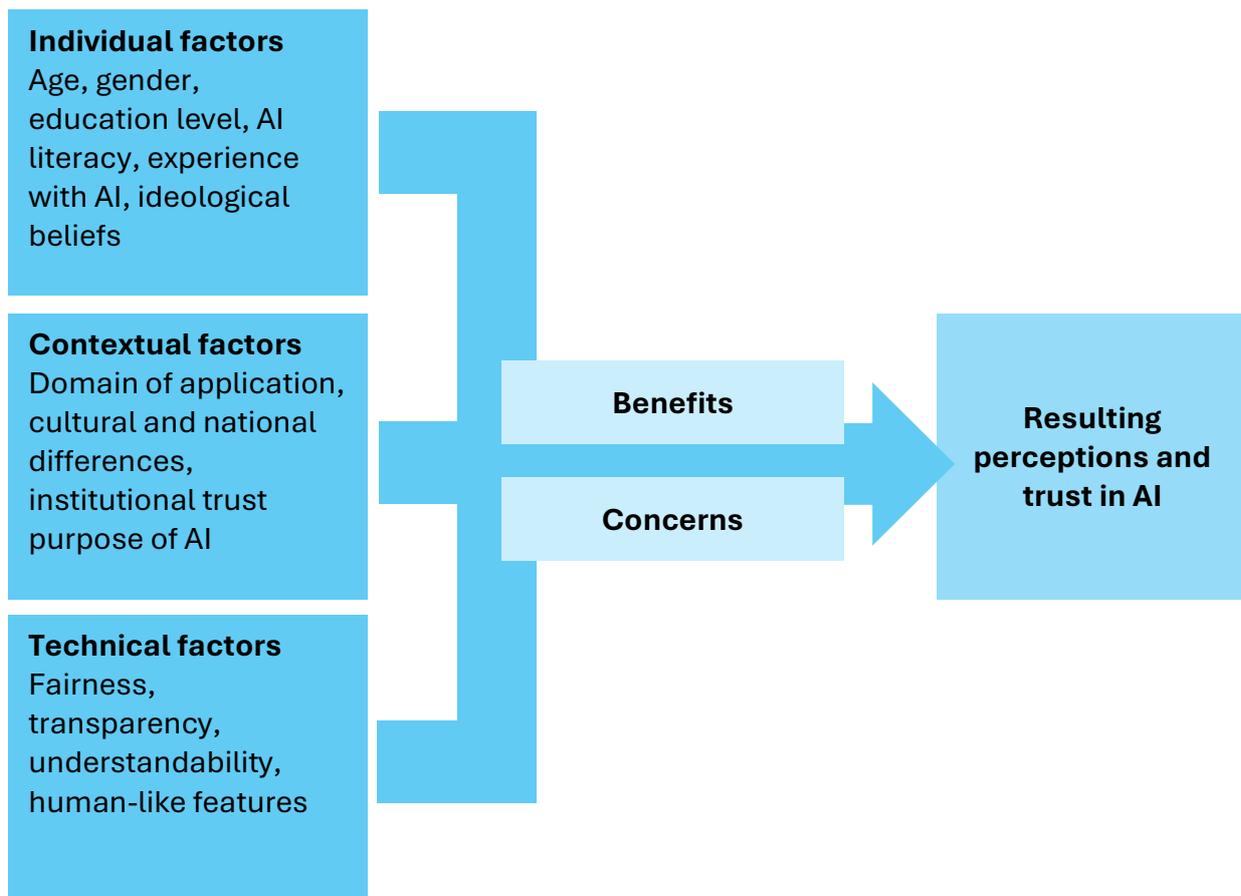

**Figure 5.** Main factors identified as relevant to perceptions and trust in AI

### Individual factors

Regardless of the domain, evidence suggest that individuals who **are younger, with higher education level, and male,** are more positive toward AI and its various applications. Individuals who are older, with lower education, and female show lower enthusiasm. This pattern can be observed across different countries (Dennis, Paz, and Yigitcanlar 2021; Kankanamge, Yigitcanlar, and Goonetilleke 2021; Mantello et al. 2024).

The interest in science and technology, as well as **practical experience or familiarity with particular applications of AI** leads to more positive attitudes towards these applications. Belonging to particular group of users (e.g., drivers or pedestrians in the context of autonomous vehicles) translates into different attitudes (Dogan, Barbier, and Peyrard 2021). It must be noted that some research suggests that familiarity with the tasks that AI automates might decrease support for AI (Horowitz et al. 2024).

The evidence on how **computer and AI literacy** influences attitudes towards AI is nuanced. Some research suggests that the increased level of computer literacy increases perception of algorithms as fair (R. Wang, Harper, and Zhu 2020) and may positively impact acceptance and adoption of AI-based solutions (Nazaretsky et al. 2022). Other studies observe that it may correlate with increased risk perceptions as it also correlates with increased awareness of the use of personal data and thus coincide with privacy concerns (Lünich, Keller, and Marcinkowski 2024).

The influence of political orientation on AI perceptions has not been studied extensively but some findings suggest that general **ideological beliefs correlate with attitudes toward AI applications**. More liberal individuals are more positive toward AI than those who are conservative (Peng 2020).

## Contextual factors

Over the last decade, AI ceased to be an abstract notion – it became essential part of the domain-specific applications. The reviewed research suggest that the **domain of application matters** and perceptions of AI significantly vary in that regard. Some research suggests that applications in the public context are more acceptable than those in the private sphere, as people across different demographics are sensitive to privacy of their data. Other research shows that the importance of factors like privacy is also context-specific, with citizens of (continental) European countries, such as Germany or France being more focused on privacy than the citizens in the United States or in South and East Asia.

**Regional, national, and cultural differences** are important. Respondents in Western and Northern countries are, in general, more concerned about harmful effects of AI than in South and East Asia, with China being consistently one of the most techno-optimistic nations (Neudert, Knuutila, and Howard 2020), at least partially due to government and media narratives (Van Noort 2024).

Several studies observed that trust in AI applications had been correlated with trust in the institutions that developed or deployed it. This mechanism is known in the literature as **trust transfer** and relates to the process by which trust established in one entity, context, or relationship is extended to another related entity or context. It has been explored particularly in the context of e-commerce and public e-services (Belanche et al. 2014; Stewart 2003). Some of the reviewed studies suggest that public trust in institutions such as the military or universities – which are usually among the most highly trusted – might translate into trust that these institutions will govern AI in a fair manner (B. Zhang and Dafoe 2020). Some of the reviewed studies suggest that public trust in institutions such as the military or universities – which are usually among the most highly trusted – might translate into trust that these institutions will govern AI in a fair manner (B. Zhang and Dafoe 2020).

**The context and intention or purpose of use matters**, even if the type of AI application remains the same. For example, the research on perception of military AI systems shows that people are more willing to accept the use of those systems in defensive rather than offensive capacity.

## Technical factors

Technical features of AI application have important effect on perceptions and trust. In particular, AI systems that pretend to be human-like are met with scepticism, with respondents preferring AI to be functional (Lehtiö et al. 2023).

Maintaining **human accountability and control of AI** is consistently one of the most important factors influencing attitudes toward applications across different domains, including autonomous vehicles (Tennant 2021), healthcare, and public administration. Overall, concerns related to the loss of human contact or decrease of humanness are among the most mentioned perceived threats of AI. It tends to matter less in the tasks that are mechanical or routine instead of human-centred (Aoki 2020; Lee 2018).

In line with AI ethics and governance literature, the reviewed studies stress importance of **reliability, fairness, transparency, and understandability** of AI systems (Shin 2021).

## 4.6. Benefits and concerns

To offer additional perspective on the results of our review, and to enable meaningful comparison of perceptions of artificial intelligence across different sectors, we propose a two-tiered categorisation of perceived benefits and concerns. This framework distinguishes between cross-domain (universal) and domain-specific themes, reflecting both shared public attitudes and context-dependent variations. Universal benefits of AI - such as improved efficiency, time-saving, error reduction, and public service enhancement - appear consistently across domains like healthcare, education, and public administration. In many cases, AI is also viewed as a marker of innovation and modernisation, especially in emerging economies. Importantly, these perceived advantages often relate to AI's role in augmenting, rather than replacing, human capacities. At the domain-specific level, more nuanced benefits are tied to particular applications: predictive diagnostics in healthcare, personalised learning in education, or early warning systems in disaster management.

Conversely, perceived concerns exhibit a similarly dual structure. Cross-domain anxieties - such as job displacement, privacy violations, and the loss of human oversight - are recurrent across all areas of AI deployment. These are often amplified by a lack of explainability and public scepticism toward the institutions deploying AI, highlighting a broader concern with institutional legitimacy and technological opacity. At the same time, domain-specific concerns reflect the ethical stakes of particular use cases. In healthcare, worries centre on misdiagnosis or dehumanisation; in autonomous vehicles, on safety and control; and in the judicial and administrative context, on fairness, bias, and the erosion of accountability. This classification enables a structured comparison of public perceptions, showing how general anxieties intersect with sectoral realities and institutional trust.

## Benefits
*Cross-domain benefits*
Across all applications of AI, people generally recognise substantial process-related benefits, particularly improvements in efficiency, innovation, resource use, and cost reduction (Gillespie et al. 2023). The perceived benefits to people themselves, such as enhancing decision-making and improving outcomes, are consistently rated lower than technical and organisational process gains. There is a marked geographic divide: populations in Brazil, India, China, South Africa (the BICS countries), and Singapore are much more likely to perceive high levels of benefit from AI, while respondents from Australia, Canada, the United Kingdom, the United States, the Netherlands, Finland, and Japan remain more sceptical. Supporting these findings, Park and Jones-Jang demonstrate that perceived ease of use (PEOU) and perceived usefulness (PU) are critical cognitive drivers of AI acceptance across sectors (Park and Jones-Jang 2023). In the American context, public support for government AI use is notably stronger when applications are linked to critical security functions, such as identifying victims of crime, underlining that perceived cross-domain benefits are tied closely to public trust and application purpose (Boudreaux et al. 2024).

*Domain-specific benefits*

People's perception of benefits varies depending on the domain of AI application, with healthcare AI being seen as offering the greatest personal benefits, particularly through improving diagnosis and treatment outcomes (Gillespie et al. 2023). In contrast, the perceived benefits of AI used in human resources, particularly for hiring and promotion, are considerably lower across many countries. This pattern holds across most regions but is especially pronounced in Western countries and Japan, where scepticism toward HR applications of AI is stronger than scepticism toward healthcare or security applications. While healthcare AI gains higher acceptance through perceived utility, financial and governmental AI applications must overcome challenges of trust related to credibility and authenticity (Park and Jones-Jang 2023). AI is viewed more favourably in domains that offer direct public value, like disaster response, but perceptions deteriorate when corporate surveillance is suspected (Yigitcanlar, Degirmenci, and Inkinen 2024). In the United States, government use of face recognition technology (FRT) garners public support primarily when confined to criminal investigations, with much less support for broader surveillance activities such as monitoring protests (Boudreaux 2024).

## Concerns

*Cross-domain concerns*

Across domains, cybersecurity risks are consistently ranked as the most significant concern globally, followed closely by fears of manipulation, privacy loss, job loss, system failure, and erosion of human rights (Gillespie et al. 2023). While concerns about bias and inaccuracy are present, they are generally rated lower than concerns about security and personal harm. Other studies confirm that security and surveillance fears undermine AI acceptance across sectors by eroding perceptions of ease and usefulness (Park and Jones-Jang 2023). Privacy breaches and unauthorised surveillance are the principal concerns among Australians regarding public sector AI (Yigitcanlar, Agdas, and Degirmenci 2023). In the United States, citizens express far greater concern about risks - such as misuse, bias, and loss of privacy - than about potential benefits when evaluating government uses of FRT, highlighting the enduring salience of cross-domain risk perceptions (Boudreaux 2024).

*Domain-specific concerns*

Specific concerns about AI also vary depending on its application, with job loss due to automation being the leading concern for AI use in India and South Africa, surpassing cybersecurity issues in importance (Gillespie et al. 2023). In Japan, the highest domain-specific fear centres around system failure, particularly given the country's heavy reliance on smart technologies. Across healthcare and security applications, people are somewhat more tolerant of risks compared to domains such as human resources and recommender systems, suggesting that perceived personal or societal benefits can moderate concern about specific AI risks. Healthcare AI faces concerns linked to security and privacy, whereas in financial and governmental uses, distrust in AI recommendations plays a larger role (Park and Jones-Jang 2023). Distinct risk sensitivity has been observed in urban AI applications, where privacy fears intensify around disaster management and service monitoring systems (Yigitcanlar, Degirmenci, and Inkinen 2024). Domain-specific concerns are highly application-dependent in the United States: while identifying criminal suspects with AI is broadly accepted, using AI for crowd surveillance at protests, voting locations, or public spaces triggers strong public opposition, emphasising that societal concerns about AI risks intensify when democratic rights are perceived to be threatened (Boudreaux 2024).

# 5. Discussion

## 5.1. Methodological, theoretical, and conceptual fragmentation

This review has revealed a notable fragmentation in how studies on public perceptions of AI approach their subject matter methodologically. While surveys are the most commonly employed tool, their designs vary considerably in terms of question framing, sampling strategies, and the contextual information provided to respondents. Some studies rely on public opinion surveys embedded in broader technology or policy research, while others are highly focused and domain-specific. Experimental designs, qualitative methods, and mixed-methods studies are present but inconsistently used and often not comparable across contexts. This lack of methodological coherence makes it difficult to synthesise findings, particularly when studies assess similar underlying constructs - such as trust or acceptance - but measure them through different instruments or under different assumptions.

Beyond methodology, there is also considerable variation in the theoretical and conceptual language used to describe how people relate to AI. Key terms such as "trust," "perception," "acceptability," "attitude," and "engagement" are frequently deployed without clear definitions, and often overlap in meaning. Few studies make explicit distinctions between emotional responses, cognitive evaluations, and behavioural intentions, and even fewer engage systematically with existing theories of technology-society interaction. As a result, similar empirical findings may be interpreted through divergent lenses, and contradictory conclusions can emerge from studies that appear to be asking the same questions. This conceptual ambiguity poses a significant challenge for policymakers and scholars attempting to draw generalisable lessons or identify robust patterns across studies. Addressing this fragmentation requires not only better integration of theoretical frameworks but also greater attention to conceptual precision and transparency in future research design.

In our review we did not attempt to collapse the differences between those varying methodologies of the reviewed papers, but we also did not propose an approach that would negotiate these methodologies. Instead, we chose to report on the results and indicate the patterns that seems to appear in the findings while acknowledging their conditionality and calling for a more systematic and robust approach for studying perceptions and trust in AI.

## 5.2. Implications for AI governance

As we have already indicated, the recent research focused on perception and especially on trust in AI is to a large extent tacitly or openly oriented toward supporting successful deployment of AI applications in different domains or at least interested in identifying barriers of adoption, with the underlying assumption that such an adoption is going to happen. While our own aims in this paper are not necessarily policy- and adoption-oriented, we might enumerate several implications for AI governance that follow from the reviewed studies.

One key implication of this review is the need to develop inclusive and targeted AI literacy initiatives. Public attitudes toward AI vary significantly across demographic groups, with older adults, women, and individuals with lower levels of formal education often expressing greater scepticism or displaying lower familiarity with the technology. Addressing these gaps requires designing AI literacy programmes that are accessible, practically oriented, and relevant to the lived experiences of different social groups. Rather than focusing solely on abstract technical knowledge, such initiatives should incorporate real-world examples and emphasise the practical applications, benefits, and limitations of AI technologies.

In parallel, efforts to build public trust in AI should leverage institutions that already command high levels of societal confidence. Universities, independent regulators, professional associations, and similar actors can play a crucial role in communicating and shaping responsible AI development. When AI applications are overseen, endorsed, or co-developed by trusted institutions, the credibility of the technologies themselves is often enhanced through what can be seen as a transfer of trust. Institutional endorsement not only legitimises AI use but also provides a framework for transparency and accountability. At the same time, the responsible approach to research and innovation that involves AI might be one of the factors influencing trustworthiness of the institutions in question. It would mean that trust transfer should be understood in a more complex and not unidirectional way and reforming AI governance in institutions could contribute to increase of trust in broader sense.

Public engagement strategies must also be responsive to the specific domain in which AI is deployed. Attitudes toward AI are not monolithic: perceptions differ markedly between sectors such as healthcare, defence, education, or transportation. For this reason, communication about AI must be tailored to domain-specific concerns - such as fairness in healthcare, safety in autonomous vehicles, or empathy in education - as well as to the cultural and regional context in which these technologies are introduced. One-size-fits-all messaging is unlikely to resonate and may risk alienating key constituencies.

Maintaining human oversight and preserving opportunities for human interaction is another essential component of trustworthy AI governance. The review suggests that public acceptance of AI is higher when systems are seen as augmenting rather than replacing human judgement, particularly in ethically sensitive or relational contexts. Designing AI systems that include clear accountability structures, visible human involvement, and meaningful avenues for human recourse can help foster reassurance and mitigate resistance. In domains such as healthcare, education, or social services, the presence of a human actor is often associated with care, empathy, and responsibility - qualities that AI systems are not perceived to replicate.

Additionally, framing AI development around public benefit and ethical commitments is central to cultivating legitimate and sustainable governance. AI initiatives should be transparently co-developed with stakeholders, grounded in normative principles such as fairness, accountability, and social good. Public communication should emphasise the intended purpose and safeguards of AI systems, distinguishing them from opaque, profit-driven deployments. Demonstrating ethical intent and a commitment to responsible data practices can enhance the perceived legitimacy of AI and promote wider societal acceptance.

Finally, effective governance must remain attentive to the diverse ways people relate to and talk about AI. Concepts such as trust, acceptance, and perceived usefulness are often used interchangeably, yet they refer to different cognitive and affective mechanisms. Trust, in particular, has multiple dimensions - ranging from interpersonal trust to systemic and institutional trust - and may not be shaped by the same factors as acceptance or use intention. Recognising these distinctions is essential when designing surveys, interpreting public attitudes, and formulating engagement strategies, ensuring that governance approaches are grounded in a nuanced understanding of how societal perceptions are formed and expressed.

### 5.3. Influence of media reporting

As part of our review, we looked at studies focused on how news media have been reporting on AI and how that reporting might influence the public. These papers are useful in capturing long-term trends in perceptions of AI, with studies focusing on both audience effects and the content and structure of AI-related reporting while showing important national and cultural differences in media frames and narratives of AI.

Longitudinal research confirms that public sentiment is influenced not only by framing but also by changes in tone over time. A 30-year content analysis of *The New York Times* (1986–2016) (Fast and Horvitz 2017) revealed a transition from largely optimistic coverage in earlier decades to more critical and ethically focused reporting in recent years, especially following breakthroughs in machine learning and high-profile public debates. Global sentiment analysis (Garvey and Maskal 2020) supports this shift, identifying a dual emphasis in AI coverage: while stories continue to promote technological advancement, references to inequality, surveillance, and bias have become increasingly prominent.

The United States, the focus of many of the reviewed studies presents mixed discursive landscape. While early US reporting often celebrated AI's potential, more recent coverage has increasingly focused on the ethical implications of automation and algorithmic bias, especially in domains such as employment and surveillance (Fast and Horvitz 2017; Garvey and Maskal 2020). These framings are also shaped by outlet ideology (Nguyen and Hekman 2024): while *The Guardian* tended to highlight concerns around justice and transparency, US outlets more frequently emphasised AI's potential for economic transformation.

A 2021 nationally representative U.S. survey experiment (Bingaman et al. 2021) showed that participants exposed to AI framed as technological progress were more supportive of related policies, while exposure to risk-oriented framings reduced support. A follow-up survey (Brewer et al. 2022) found that support for AI was also shaped by trust in media and interpersonal conversations, with emotions such as fear and hope mediating this relationship. A 2024 survey in the US (Owsley and Greenwood 2024) found that while most respondents had heard of AI, only a small proportion reported understanding how it worked, with those who felt more knowledgeable also expressing more positive views.

Another study (Nader et al. 2024) confirmed that lower comprehension levels were associated with increased susceptibility to negative media framings and decreased support for AI adoption in both public and private domains.

These trends are further complicated by the emotional dimension of media narratives. A 2024 US survey study (Nah et al. 2024) showed that AI news use could simultaneously generate trust and fear, with trust moderating the perceived benefits of AI for society. Editorial positioning and section placement within news outlets also contribute to public perception. An analysis of 399 articles from *The New York Times*, *The Wall Street Journal*, and *The Washington Post* (Chuan, Tsai, and Cho 2019) revealed that AI stories were largely situated in business and technology sections, where the dominant frame was one of productivity and profit. Ethical concerns, where present, appeared mostly in editorials or features rather than front-page reporting. A comparative framing analysis of *The New York Times* and *The Guardian* (Nguyen and Hekman 2024) added that left-leaning newspapers were more likely to focus on fairness and inequality, while centre-right sources leaned towards narratives of competitiveness and national leadership.

In Europe, media and political discourses tend to reflect both techno-optimism and democratic scepticism. A 2022 study of German news and parliamentary documents (Köstler and Ossewaarde 2022) characterised AI as a driver of national development, but also noted tensions around inclusion, accountability, and regulatory oversight. These debates often intersect with industrial and educational policy, positioning AI as both an opportunity and a policy challenge.

Survey data from across Europe confirm that these differences in media framing are linked to divergent public perceptions. A 2023 cross-national study (Vorobeva et al. 2024) found that respondents in Northern Europe, including Finland and Norway, were more likely to associate AI with innovation and convenience, while those in Southern Europe, such as Portugal and Italy, expressed higher levels of concern about risk, fairness, and institutional trust. These regional distinctions reflect not only different media ecosystems but also broader cultural and political expectations surrounding technological change.

National and cultural differences in reporting are particularly evident when comparing Western and non-Western media ecosystems. A sentiment analysis of 2,240 AI-related articles published between 2016 and 2021 in *Süddeutsche Zeitung*, *Die Welt*, and *Die Zeit* (Rana et al. 2024) found that 86% of German articles framed AI positively, with a strong emphasis on scientific innovation and economic modernisation. A complementary discourse analysis of 178 articles from German newspapers and magazines between 2015 and 2021 (Carstensen and Ganz 2023) found that portrayals of AI leadership were overwhelmingly male-coded, reinforcing existing gendered narratives within technology reporting. In contrast, a content analysis of Chinese state-aligned media (Van Noort 2024) found a consistent emphasis on emotional positivity, particularly national pride and optimism, while risk-related framings were almost entirely absent. These results illustrate the selective emotional framing used by different media systems to shape public sentiment.

## 5.4. Limitations of the study

While this systematic literature review offers a comprehensive overview of existing studies on public perceptions of artificial intelligence, several limitations should be acknowledged.

First, the diversity of research questions and methodological approaches limits the generalisability of the findings. Although the majority of reviewed items rely on similar methods - primarily surveys - they are situated in varied national and cultural contexts, asking different questions at different points in time. These variations make direct comparisons across studies difficult. Consequently, we have avoided overly formalising our analysis, opting instead to identify consistent patterns and illustrate them through reference to specific studies.

Second, the sample of reviewed literature is affected by publication bias, which may skew the findings toward particular geographical regions and domains of application. The majority of studies originate from the United States, Europe, Australia, and, to a lesser extent, countries in South and East Asia. In contrast, research from South America and Africa is severely underrepresented. This imbalance illuminates a broader issue in technology and social science research and severely limits our ability to draw global conclusions about public attitudes toward AI.

Third, conceptual ambiguity surrounding key terms such as "trust" and "perception" complicates interpretation. Most studies do not define these terms with precision, and their meanings vary not only across academic traditions but also across cultural and linguistic contexts. This conceptual vagueness contributes to inconsistencies in how findings are reported and understood, further constraining the generalisability of the results.

Fourth, there exists a temporal gap between technological developments in AI and the research that seeks to assess public responses to them. Recent advances - such as the release of publicly available, multimodal generative AI tools (e.g., text-to-video models) in the latter half of 2024 and early 2025 - are not captured in the reviewed literature. Likewise, shifts in public discourse surrounding AI, such as those linked to the global election cycle of 2024 and the new U.S. presidential administration, are likely to shape perceptions in ways that will only be reflected in future publications.

Fifth, the reviewed literature offers limited comparative analysis between expert and public perceptions of AI. While most studies focus on general public responses, certain domains - particularly education and healthcare - include research based on expert opinions. Notably, these "experts" are often domain specialists (e.g., teachers or medical practitioners) rather than AI professionals. Only a few studies attempted to directly compare expert and public views within the same research design. However, there is some indication that domain experts' attitudes, like those of the general public, are shaped by familiarity and experience with AI systems. For instance, educators and healthcare professionals with greater exposure to AI tend to express more favourable views on its integration into their work.

Sixth, our review did not systematically include studies focused on public perceptions of future-oriented or speculative AI, such as Artificial General Intelligence (AGI) or Artificial Superintelligence (ASI). Although the technical feasibility of such systems remains contested, they constitute a central element of the public imagination and media discourse around AI. As such, they may significantly influence attitudes toward existing technologies, despite not being the subject of focused empirical research in the reviewed literature.

Seventh, the scope of this review was limited to literature published in English. As a result, studies written in other languages - potentially offering valuable insights into national and cultural variations in AI perception - were excluded. Future research aiming to capture more nuanced, country-specific understandings of AI should address this limitation by incorporating non-English sources.

Finally, the review focused exclusively on perceptions of AI rather than on how individuals actually engage with AI technologies in practice. While the included studies provide insight into opinions, attitudes, and imagined futures, they do not explore in detail how AI is used in everyday life or professional settings. A future line of research could complement this literature by examining lived experiences and practices surrounding AI adoption and use across different social and institutional contexts.

## 6. Conclusions

This review confirms that the growing body of research on public perceptions of artificial intelligence has emerged primarily in response to the perceived need for understanding how individuals and communities orient themselves toward AI across a range of domains and use cases. Much of this literature is teleological in nature - seeking to identify perceived benefits and risks, uncover drivers of trust or distrust, and assess levels of acceptance or resistance to AI systems. These efforts are often framed around policy-relevant questions about how to govern, regulate, or communicate AI innovations to the public.

However, our analysis also reveals that the field remains conceptually and methodologically fragmented. The studies we reviewed employ a wide array of theoretical lenses - ranging from psychological models of technology acceptance to sociotechnical or institutional approaches - but without consistent engagement across paradigms. Likewise, there is little agreement on how key constructs such as "trust," "acceptability," or even "public" are defined and operationalised. As a result, comparing findings across studies or synthesising general insights into public attitudes remains challenging.

We argue that future research - including the study we are currently developing - should aim to address this fragmentation by critically mapping and synthesising the diverse frameworks that underpin the field. Such work would not only enhance conceptual clarity but also enable more meaningful comparisons across different contexts and technologies.

Finally, we note a pronounced geographical imbalance in the current literature. The overwhelming majority of empirical studies focus on public attitudes in North America, Europe, and other high-income countries. This replicates a broader pattern observed in technology studies and survey research more generally, where perspectives from the Global South remain severely underrepresented. Addressing this gap is crucial - not only for reasons of equity and inclusion but also because AI systems are increasingly deployed globally, often in vastly different institutional and cultural settings. Understanding how AI is perceived and received in these contexts will be essential to the development of just and context-sensitive governance frameworks.

# Acknowledgement

This research was funded by the North West Partnership for Security and Trust, which is funded through GCHQ. The funding arrangements required this paper to be reviewed to ensure that its contents did not violate the Official Secrets Act nor disclose sensitive, classified and/or personal information.